\documentclass[10pt]{amsart}

\usepackage[T1]{fontenc} 
\usepackage[utf8]{inputenc} 
\usepackage{longtable}
\usepackage{amssymb}
\usepackage{amsthm}
\usepackage{amsmath}
\usepackage{wasysym}
\usepackage{dcpic, pictex}
\usepackage{bbm}
\usepackage{enumerate}
\usepackage{amsfonts}
\usepackage{amsthm}
\usepackage{amsmath}
\usepackage{times}
\usepackage{amscd}
\usepackage[mathscr]{eucal}
\usepackage{indentfirst}
\usepackage{pict2e}
\usepackage{epic}
\usepackage{epstopdf} 
\usepackage{verbatim}
\usepackage{cancel}
\usepackage[foot]{amsaddr}
\usepackage{lipsum}
\usepackage{mathrsfs}
\usepackage{bm}

\usepackage{braket}

\usepackage{tikz}
\usetikzlibrary{decorations.pathreplacing,decorations.markings}
\tikzset{
	on each segment/.style={
		decorate,
		decoration={
			show path construction,
			moveto code={},
			lineto code={
				\path [#1]
				(\tikzinputsegmentfirst) -- (\tikzinputsegmentlast);
			},
			curveto code={
				\path [#1] (\tikzinputsegmentfirst)
				.. controls
				(\tikzinputsegmentsupporta) and (\tikzinputsegmentsupportb)
				..
				(\tikzinputsegmentlast);
			},
			closepath code={
				\path [#1]
				(\tikzinputsegmentfirst) -- (\tikzinputsegmentlast);
			},
		},
	},
	mid arrow/.style={postaction={decorate,decoration={
				markings,
				mark=at position .5 with {\arrow[#1]{stealth}}
	}}},
}

\usepackage{
amsfonts, 
latexsym, 
enumerate,
}
\usepackage[english]{babel}
\allowdisplaybreaks
\makeindex
\newtheorem{theorem}{Theorem}[section]
\newtheorem{lemma}[theorem]{Lemma}
\newtheorem{proposition}[theorem]{Proposition}
\newtheorem{definition}[theorem]{Definition}

\newtheorem{remark}[theorem]{Remark}
\newtheorem{corollary}[theorem]{Corollary}
\newtheorem{assumption}[theorem]{Assumption}

\usepackage[a4paper,top=3.5cm,bottom=3.5cm,left=3cm,right=3cm]{geometry}
\numberwithin{equation}{section}
\allowdisplaybreaks

\usepackage{ bbold }

\usepackage{xcolor}

\newcommand{\nc}{\normalcolor}
\newcommand{\dif}{\mathrm{d}}
\newcommand{\E}{\mathbf{E}}
\newcommand{\R}{\mathbf{R}}
\newcommand{\C}{\mathbf{C}}

\newcommand{\N}{\mathbf{N}}

\newcommand{\ii}{\mathrm{i}}
\newcommand{\ee}{\mathrm{e}}
\newcommand{\rd}{\mathrm{d}}

\newcommand{\ms}{\mathfrak{s}}
\newcommand{\mc}{\mathfrak{c}}
\newcommand{\mb}{\mathfrak{b}}

\newcommand{\Los}{\mathfrak{M}}
\newcommand{\Losa}{\overline{\mathfrak{M}}}
\newcommand{\Epa}{\overline{\mathfrak{P}}}
\newcommand{\Over}{\mathfrak{S}}
\newcommand{\Overa}{\overline{\mathfrak{S}}}

\usepackage{hyperref}

\title{Loschmidt echo for deformed Wigner matrices}

\thanks{All authors were supported by the ERC Advanced Grant ``RMTBeyond'' No.~101020331. }

\date{\today}

\begin{document}

	\maketitle
\vspace{0.25cm}

\renewcommand{\thefootnote}{\fnsymbol{footnote}}

\noindent
\mbox{}%
\hfill%
\begin{minipage}{0.21\textwidth}
	\centering
	{L\'aszl\'o Erd\H{o}s}\footnotemark[1]\\
	\footnotesize{\textit{lerdos@ist.ac.at}}
\end{minipage}
\hfill%
\begin{minipage}{0.21\textwidth}
	\centering
	{Joscha Henheik}\footnotemark[1]\\
	\footnotesize{\textit{joscha.henheik@ist.ac.at}}
\end{minipage}
\hfill%
\begin{minipage}{0.21\textwidth}
	\centering
	{Oleksii Kolupaiev}\footnotemark[1]\\
	\footnotesize{\textit{oleksii.kolupaiev@ist.ac.at}}
\end{minipage}
\hfill%
\mbox{}%
\footnotetext[1]{Institute of Science and Technology Austria, Am Campus 1, 3400 Klosterneuburg, Austria. 
}

\renewcommand*{\thefootnote}{\arabic{footnote}}
\vspace{0.25cm}

\begin{abstract} 
We consider two Hamiltonians that are close to each other, $H_1 \approx H_2 $, and analyze the time-decay of the corresponding \emph{Loschmidt echo} $\mathfrak{M}(t) := |\langle \psi_0, \mathrm{e}^{\mathrm{i} t H_2} \mathrm{e}^{-\mathrm{i} t H_1} \psi_0 \rangle|^2$ that expresses the effect of an imperfect time reversal
	 on the initial  state $\psi_0$. 
Our model Hamiltonians are deformed Wigner matrices that do not share a common eigenbasis. The main 
tools for our results are two-resolvent laws  for such  $H_1$ and $H_2$. 	
\end{abstract}
\vspace{0.15cm}

\footnotesize \textit{Keywords:} Quantum Dynamics, Loschmidt echo, Matrix Dyson Equation.

\footnotesize \textit{2020 Mathematics Subject Classification:} 60B20, 82C10.
\vspace{0.25cm}
\normalsize

\section{Introduction}
\label{sec:introduction}

 Recent quantum technological advances put quantum mechanical time reversal procedures in the focus of both experimental \cite{Hahn50, LevUnaPastJCP98, Linnemann16, Sanchez16, Gaerttner17, Niknam20} and theoretical \cite{Haakebook, 0806.0987, 0904.0172, SchmittKehrein16, SerbynAbanin, SchmittKehrein1711.00015, DRecho20, DRecho20ZF} research (see also the review \cite{scholarpedia} for a concise overview). The basic physical setup consists of an initial (normalized) quantum state $\psi_0$ and two self-adjoint Hamiltonians close to each other,   $H_1 \approx H_2$, each governing the evolution of the system during a time span $t$. First, the initial state $\psi_0$ evolves under the Hamiltonian $H_1$ from time zero to $t$, resulting in the state $\psi_t = \exp{(-\ii  H_1 t)} \psi_0$. Then, during a second evolution between $t$ and $2t$, one applies the Hamiltonian
 $H_2$ backward in time, equivalently the  Hamiltonian $-H_2$ in forward time,  aiming to recover the initial state $\psi_0$.  A schematic summary of this process is given by
 \begin{equation} \label{eq:summary1}
 	\psi_0 \xrightarrow[~~~H_1~~~]{t} \psi_t \xrightarrow[~~-H_2~~]{t} \psi_0' \,. 
 \end{equation}
 Note that, if $H_2 = H_1$, the restoration of $\psi_0$ would be perfect, $\psi_0’ =\psi_0$ for any time $t$. 
 However, in realistic setup the second Hamiltonian
  is never a perfect copy of the first one and  the non-zero difference between $H_1$ and $H_2$ regularly
   leads to an imperfect recovery  $\psi_0'$ of $\psi_0$ and the discrepancy also depends on time.

 This imperfection in the time reversal is captured in the scalar \emph{overlap function} \cite{0410202, GPSZ06, SerbynAbanin} (sometimes
 also called \emph{fidelity amplitude}~\cite{0311022, GPSZ06, 1510.03140})
\begin{equation} \label{eq:overintro}
\Over(t)  = \Over_{H_1, H_2}^{(E_0)}(t) := \big\langle \psi_0, \ee^{\ii H_2 t} \ee^{-\ii H_1 t} \psi_0 \big\rangle
\end{equation}
where it is  assumed that
 the initial state is supported\footnote{This means that, when writing $\psi_0 = \sum_n c_n^{(i)} \phi_n^{(i)}$ in the eigenbasis $\{\phi_n^{(i)}\}_n$ of $H_i$, only coefficients $c_n^{(i)}$ corresponding to an eigenvalue close to $E_0$ are non-vanishing.} around its energy $\langle \psi_0, H_1 \psi_0 \rangle \approx \langle \psi_0, H_2 \psi_0 \rangle \approx E_0$. The central object of our paper is the absolute value square of the overlap function
\begin{equation} \label{eq:echo1intro}
	\Los(t) = \Los_{H_1, H_2}^{(E_0)}(t) := \left|\Over_{H_1, H_2}^{(E_0)}(t)  \right|^2\,.
\end{equation}
This was coined the \emph{fidelity}, e.g.,  by Gorin \emph{et.~al.} in \cite{GPSZ06}, or
the \emph{Loschmidt echo} by Peres \cite{peres} and Jalabert-Pastawski in \cite{JalaPasPRL}
owing to its connection to the classical Loschmidt's paradox of time reversibility \cite{Loschmidt1876, Boltzmann1877}.
 
In addition to \eqref{eq:overintro}--\eqref{eq:echo1intro}, we will also consider an \emph{averaged overlap function} and an \emph{averaged Loschmidt echo}, defined as
\begin{equation} \label{eq:OverLosav}
	\Overa(t) = \Overa_{H_1, H_2}^{(E_0, \eta_0)}(t) := \mathrm{Av}\big[ \Over_{H_1, H_2}^{(E)}(t)\big] \quad \text{and} \quad 	\Losa(t)  = \Losa_{H_1, H_2}^{(E_0, \eta_0)}(t):= \left|\Overa_{H_1, H_2}^{(E_0, \eta_0 )}(t)  \right|^2,
\end{equation}
respectively. In \eqref{eq:OverLosav}, by $\mathrm{Av}[...]$, we denoted an averaging over 
initial states with energies $E$ in a small energy window of size   $\eta_0$ around
 $E_0$ (see \eqref{eq:locechodef} below for a precise implementation of this). 

\bigskip 

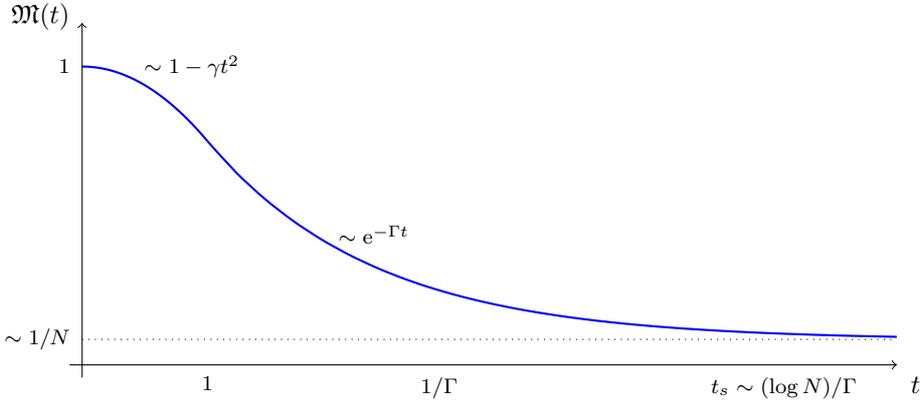
\begin{figure}[h]
	\centering
			\begin{tikzpicture}[scale=\textwidth/18.2cm]
	\draw[->] (-0.2,0) -- (13,0);
	\draw (13.3,0) node[below=1pt] {$t$};
	\draw[->] (0, -0.2) -- (0,5.5);
	\draw (0,5.6) node[left=1pt] {$\mathfrak{M}(t)$};
	\draw (0,4.8) node[left=1pt] {\footnotesize $1$};

	\draw[dotted] (0,0.41) -- (13,0.41);
	\draw (0,0.41) node[left=1pt] {\footnotesize $\sim 1/N$};
	
	\draw (2,0) node[below=1pt] {\footnotesize$1$};
	\draw (5.7,0) node[below=1pt] {\footnotesize$1/\Gamma$};
	\draw (11.2,0) node[below=1pt] {\footnotesize$t_s \sim  (\log N)/\Gamma$};

	\draw (0.8,4.8) node[right=1pt] {\footnotesize $\sim 1 - \gamma t^2$};
	\draw (3.9,2.1) node[right=1pt] {\footnotesize $\sim \ee^{-\Gamma t}$};

	\draw[blue, thick, smooth] plot[variable=\x, domain=0:2] (\x, {4.8-0.3*\x^2});
	\draw[blue,thick,smooth] plot[variable=\x,domain=2:13] (\x,{0.4+3.2*exp(-1.2*(\x-2)/3.2)});
	
\end{tikzpicture}
\caption{Illustrated is the typical behavior of the Loschmidt echo in its three consecutive phases: Short-time parabolic decay, intermediate-time asymptotic decay, and long-time saturation. In both of our main results \eqref{eq:firstthm}--\eqref{eq:secondthm}, the decay parameters $\gamma$ and $\Gamma$ generally satisfy $\gamma \sim \Gamma \sim \langle (H_1 - H_2)^2\rangle$; cf.~\eqref{eq:gammaGammageneral}.}
\label{fig:1}
\end{figure}

The Loschmidt echo is a basic object in the study of complex quantum system and has attracted considerable attention in different areas of research, e.g.~quantum chaos \cite{peres, JalaPasPRL, 0904.0172, Haakebook, 0806.0987, SerbynAbanin}, quantum information theory \cite{0105149, 0809.4416}, and statistical mechanics \cite{SchmittKehrein16, SchmittKehrein1711.00015, DRecho20, DRecho20ZF}.
The Loschmidt echo, as a measurable physical quantity, is  observed and predicted to follow
a quite universal behavior as a function of time (cf.~the discussion of our main results around \eqref{eq:firstthm}--\eqref{eq:secondthm} below). On a high level (see \cite{scholarpedia}), the reason for the  robust universal features
 is that the subsequent forward and backward evolutions act as a ``filter" for irrelevant details.
 The typical behavior of the Loschmidt echo can be structured in three consecutive phases 
 (see Figure~\ref{fig:1}, cf.~also \cite[Figure~4]{scholarpedia}): 
 After an initial short-time parabolic decay, $\Los(t) \approx 1 - \gamma t^2$, the Loschmidt echo exhibits an intermediate-time asymptotic  exponential \nc 
 decay\footnote{In the very extreme case, when the difference $H_1 - H_2$ is small compared to the local eigenvalue spacing one observes Gaussian instead of exponential decay,  $\Los(t) \approx \ee^{-\gamma t^2}$ (see, e.g., \cite[Section~2.3.1]{scholarpedia})}, 
 $\Los(t) \approx \ee^{-\Gamma t}$. 
 Finally, at times $t$ beyond the so-called \emph{saturation time} $t_s \sim  (\log N)/\Gamma$, 
 where $N$ is the (effective) Hilbert space dimension, it 
 saturates at a value inversely proportional to $N$, i.e.  $\Los(t) \sim 1/N$. 

There are several ways to determine the behavior of the Loschmidt echo in a given system (see the review \cite{scholarpedia}): One standard option is to employ semi-classical approximations \cite{JalaPasPRL, 0410202, 1106.4027}, another one is numerical evaluation \cite{TalEzer, DeRaedt, vanDijk}.  
Here, following E.~Wigner's original vision of describing  chaotic quantum systems by large random matrices \cite{Wigner} 
(see also further extensive physics literature \cite{0112015, 0311022, Kohler, DRecho20, DRecho20ZF}), 
we model (part of) the Hamiltonian(s) $H_1, H_2$ by Wigner random matrices with independent entries.
In this setup we can give a mathematically rigorous and quite precise analysis of certain features of
the Loschmidt echo; some of them have been predicted in the physics literature.   

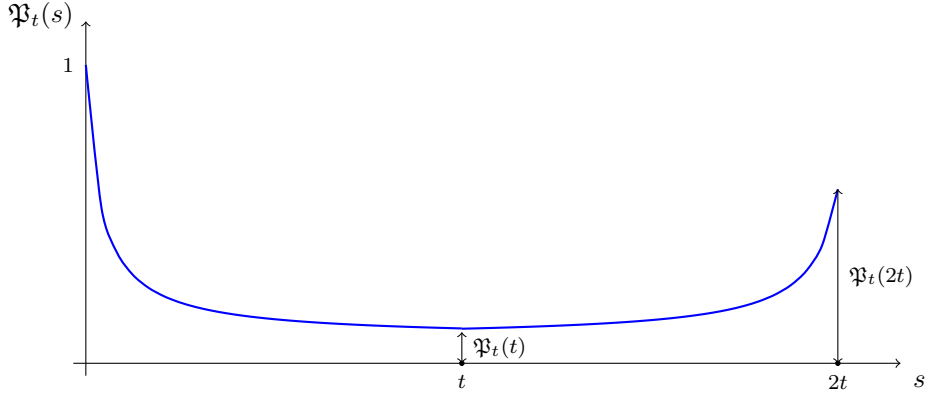
\begin{figure}[h]
	\centering
\begin{tikzpicture}[scale=\textwidth/18.2cm]
	\draw[->] (-0.2,0) -- (13,0);
	\draw (13.3,0) node[below=1pt] {$s$};
	\draw[->] (0, -0.2) -- (0,5.5);
	\draw (0,5.6) node[left=1pt] {$\mathfrak{P}_t(s)$};
	\draw (0,4.8) node[left=1pt] {\footnotesize $1$};
	\draw (6,0) node[below=1pt] {\footnotesize $t$};
    \filldraw [black] (6,0) circle (1pt);
	\draw (12,0) node[below=1pt] {\footnotesize $2t$};
    \filldraw [black] (12,0) circle (1pt);
    	
	\draw[blue, thick, smooth] plot[variable=\x, domain=0:6] (\x, {0.4+1/(\x+1/4.4)});
	\draw[blue, thick, smooth] plot[variable=\x, domain=6:12] (\x, {0.4+1/(12-\x+1/2.4)});
	
	\draw[<->] (6,0) -- (6,0.5);
	\draw (6,0.25) node[right=1pt] {\footnotesize $\mathfrak{P}_t(t)$};
	\draw[<->] (12,0) -- (12,2.8);
	\draw (12,1.4) node[right=1pt] {\footnotesize $\mathfrak{P}_t(2t)$};
\end{tikzpicture}
\caption{Schematic behavior of the overlap $\mathfrak{P}_t(s)$ from \eqref{eq:def_P} for $s\in [0,2t]$. At the midpoint, $s=t$, typically $\mathfrak{P}_t(t) \ll \mathfrak{P}_t(2t)$, which indicates a partial recovery between time $t$ and $2t$ of the original complete overlap at time $s=0$.}
\label{fig:2}
\end{figure}

 Before defining the precise model,  we first \nc discuss where the name \emph{echo} for $\mathfrak{M}(t)$ comes from. Fix any time $t>0$ and consider the two-step process \eqref{eq:summary1}. For $s\in [0,2t]$ denote the state at the intermediate time $s$ by $\psi_s$, namely, $\psi_s = \ee^{-\ii sH_1}\psi_0$ for $s\in[0,t]$ and $\psi_s=\ee^{\ii (t-s)H_2}\ee^{-\ii t H_1}\psi_0$ for $s\in[t,2t]$. Comparing this notation to \eqref{eq:summary1} we see that $\psi_{2t}=\psi_0'$. Denote further the (squared) overlap of $\psi_0$ and $\psi_s$ by 
\begin{equation}
\mathfrak{P}_t(s):=\vert \langle\psi_0,\psi_s\rangle\vert^2.
\label{eq:def_P}
\end{equation}
This quantity depends also on $\psi_0$ and $H_1, H_2$, but we suppress this dependence in notations for simplicity. We call $\mathfrak{P}_t(s)$, $s\in[0,2t]$, the \emph{Loschmidt echo process}. Clearly, $\mathfrak{P}_t(0)=1$ and $\mathfrak{P}_t(2t)=\mathfrak{M}(t)$. Later in Corollary \ref{cor:2bumps} we show that $\Epa_t(t)\ll \Epa_t(2t)$ under suitable assumptions, where $\Epa_t$ is an averaged version of $\mathfrak{P}_t$ defined in \eqref{eq:Av_Pr1}-\eqref{eq:Av_Pr2}. This result means that typically the original complete overlap $\mathfrak{P}_t(0)=1$ is partially recovered at the final moment of time $2t$, though at the intermediate time $t$ it is much smaller than $\mathfrak{P}_t(2t)$ (see Figure~\ref{fig:2}).

As our main result, we rigorously prove the decay of the Loschmidt echo for two different physical settings 
(called \emph{Scenario I} and \emph{Scenario II}), which we now describe 
somewhat informally (see Section~\ref{sec:mainres} for more precise statements containing all the technical details). 

\bigskip 

 For our first result (Scenario I, Theorem \ref{thm:echo1}), we consider two \emph{deformed Wigner matrices}
  $H_j = D_j + W$, $j \in [2]$, with bounded deterministic $D_j$, satisfying 
  $D_1 \approx D_2$, and $W$ a (common) random Wigner matrix. 
  This setup
 corresponds to an arbitrary deterministic system modeled by the Hamiltonian $D_1$ 
 and the time reversed Hamiltonian $D_2$ nearby, which are both 
 subject to an overall mean-field noise described by the same
Wigner matrix $W$ throughout the whole echo process. In this setting, for an energy $E_0$ in  the bulk of the density of states of both $H_1$ and $H_2$, we consider the averaged Loschmidt echo \eqref{eq:OverLosav}. 
Our result in Theorem \ref{thm:echo1} then shows (i) \emph{short-time parabolic decay} and (ii) \emph{intermediate-time asymptotic decay} of the form 
\begin{equation} \label{eq:firstthm}
	\Losa(t) \approx \begin{cases}
		1 - \gamma t^2 \quad &\text{for} \quad t \ll 1  \\
		\mathrm{e}^{- \Gamma t} \quad &\text{for} \quad 1 \ll t \lesssim \Delta^{-2}.
	\end{cases}
\end{equation}
Both \emph{decay parameters} satisfy $\gamma \sim \Delta^2$ and $\Gamma \sim \Delta^2$, where $\Delta := \langle (D_1 -D_2)^2 \rangle^{1/2}$, and depend on $E_0$ and the density of states at $E_0$. The quadratic relation $\Gamma \sim \Delta^2$ is in perfect agreement with \emph{Fermi's golden rule}. 

For our second main result (Scenario II, Theorem \ref{thm:echo2}), we consider a physically different situation: Now the two Hamiltonians\footnote{Within Theorem \ref{thm:echo2}, the Hamiltonians $H_1 = D$ and $H_2 = D + \lambda W$ will be denoted by $H_0$ and $H_\lambda$, respectively.} are $H_1 = D$ and $H_2 = D + \lambda W$ with a deterministic $D$, a Wigner matrix $W$ and a small parameter $|\lambda| \ll 1$. Hence, the imperfection along the backward evolution is modeled by a small Wigner matrix $\lambda W$ indicating an additive noise 
 (see, e.g., \cite[Eq.~(31)]{DRecho20} or \cite[Eq.~(1)]{DRecho20ZF}). For a normalized initial state $\psi_0 \in \C^N$ supported in the bulk of the density of states of both $H_1$ and $H_2$ with energy $\langle \psi_0, H_1 \psi_0 \rangle \approx \langle \psi_0, H_2 \psi_0 \rangle \approx E_0$, we now consider the usual Loschmidt echo \eqref{eq:echo1intro} without averaging. 
 Similarly to \eqref{eq:firstthm}, our result in Theorem \ref{thm:echo2} then shows (i) \emph{short-time parabolic decay} and (ii) \emph{intermediate-time asymptotic decay}  of the form
\begin{equation} \label{eq:secondthm}
	\Los(t) \approx \begin{cases}
		1 - \gamma t^2 \quad &\text{for} \quad t \ll 1  \\
		\mathrm{e}^{- \Gamma t} \quad &\text{for} \quad 1 \ll t \lesssim \lambda^{-2}.
	\end{cases}
\end{equation}
Here the \emph{decay parameters} satisfy $\gamma = \lambda^2$ and $\Gamma = 2 \pi \rho_0(E_0) \lambda^2$, where $\rho_0$ is the (limiting) density of states of $D$. 
 Finally, we note that, since  $\E \langle W^2 \rangle = 1$, in both of our results \eqref{eq:firstthm}--\eqref{eq:secondthm} the decay parameters $\gamma$ and $\Gamma$ satisfy the general relation
\begin{equation} \label{eq:gammaGammageneral}
\gamma \sim \Gamma \sim \E \langle (H_1 -H_2)^2 \rangle\,. 
\end{equation}

As corollaries to our main results  \eqref{eq:firstthm}--\eqref{eq:secondthm} in Theorems \ref{thm:echo1} and \ref{thm:echo2}, we also consider the \emph{scrambled Loschmidt echo} \cite{DRecho20, KurchanJSP2018, SchmittKehrein1711.00015} $\Los_\delta^{\mathrm sc}(t)$ and its averaged analog $\Losa_\delta^{\mathrm sc}(t)$.  They are obtained from 
\begin{equation} \label{eq:overintroSC}
	\Over_\delta^{\mathrm{sc}}(t)   := \big\langle \psi_0, \ee^{\ii H_2 t} \ee^{-\ii \delta V}\ee^{-\ii H_1 t} \psi_0 \big\rangle
\end{equation}
and its averaged analog $\Overa_\delta^{\mathrm{sc}}(t)$ as
$$
  \Los_\delta^{\mathrm sc}(t): = \big| \Over_\delta^{\mathrm{sc}}(t)  \big|^2 \quad \text{and} \quad 
   \Losa_\delta^{\mathrm sc}(t): = \big| \Overa_\delta^{\mathrm{sc}}(t)  \big|^2,
  $$
 exactly as in \eqref{eq:echo1intro}--\eqref{eq:OverLosav}, respectively.  In \eqref{eq:overintroSC},  $H_1$ and $H_2$ are the two Hamiltonians either from Scenario I or Scenario II. 
 The idea behind the quantity in \eqref{eq:overintroSC} is that, between the forward and backward evolution, there is a (short) \emph{scrambling time} $\delta$, in which the system is uncontrolled and governed by another self-adjoint \emph{scrambling Hamiltonian} $V$ \cite{DRecho20}. Similarly to \eqref{eq:summary1}, a schematic summary of this process is given by
\begin{equation} \label{eq:summary2}
	\psi_0 \xrightarrow[~~~H_1~~~]{t} \psi_t \xrightarrow[~~~V~~~]{\delta} \psi_t' \xrightarrow[~~-H_2~~]{t} \psi_0' \,. 
\end{equation}
In Corollaries \ref{cor:echo1} and \ref{cor:echo2} (of Theorems \ref{thm:echo1} and \ref{thm:echo2}, respectively), we model the scrambling Hamiltonian by another Wigner matrix, $V := \widetilde{W}$, that is \emph{independent} of $W$; see \cite{DRecho20}. As a result, we find that 
\begin{equation} \label{eq:corollaries}
\Losa_\delta^{\mathrm sc}(t) \approx (\varphi(\delta))^2 \, \Losa(t) \quad \text{and} \quad \Los_\delta^{\mathrm sc}(t) \approx (\varphi(\delta))^2 \, \Los(t) 
\end{equation}
in the setting of Scenario I and Scenario II, respectively,  where we denoted $\varphi(\delta)  :=  J_1(2 \delta)/\delta$ and $J_1$ is the first order Bessel function of the first kind. Note that in \eqref{eq:corollaries} we see the effects of the scrambling Hamiltonian $V$ and the imperfect time reversal of $H_1$ and $H_2$ to completely decouple (cf.~\cite[Eq.~(35)]{DRecho20}). 

 We point out that Scenario II, discussed around \eqref{eq:secondthm}, and the corollaries described in \eqref{eq:corollaries} are primarily given to provide a more
comprehensive view of Loschmidt echoes modeled with Wigner matrices. Technically, these are obtained by simple modifications of earlier results and techniques \cite{pretherm, thermalization} (see the proof in Section \ref{sec:pfecho23} for details). The mathematically novel  principal part of this work therefore focuses on Theorem \ref{thm:echo1}
analyzing Scenario I. 

\bigskip 

The proof of Theorem \ref{thm:echo1} relies on a \emph{two-resolvent global law}, i.e.~a concentration estimate for products of resolvents $G_i(z_i) := (H_i - z_i)^{-1}$ for $z_i \in \C\setminus \R$ as the dimension $N$ of the matrix becomes large. 
 By functional calculus, this can then be used for
 computing more complicated functions of $H_i$, like the exponential, and thus connecting to the time evolutions above. A typical global law computes, e.g., 
\begin{equation} \label{eq:globallawintro}
	\langle \psi_0, G_2(z_2) G_1(z_1) \psi_0 \rangle
\end{equation}
to leading order in $N$ with error terms vanishing like $N^{-1/2+\epsilon}$ with very high probability. Note that such an error term prohibits accessing times beyond the saturation time (when the Loschmidt echo is of order $1/N$), which is why our main results \eqref{eq:firstthm}--\eqref{eq:secondthm} are only valid up to order $\Delta^{-2}$ and $\lambda^{-2}$, respectively, i.e.~well before the saturation time. 

The main novelty of this paper is a precise estimate on the deterministic leading term to \eqref{eq:globallawintro}. While $G_i(z_i) \approx M_i(z_i)$, where the deterministic matrix $M_i(z_i)$ is the solution of the \emph{Matrix Dyson Equation} \eqref{eq:MDEfirst}, it does \emph{not} hold that $G_2(z_2) G_1(z_1) \approx M_2(z_2) M_1(z_1)$ owing
to correlations between $G_1$ and $G_2$.  The correct approximation is  
\begin{equation} \label{eq:m12}
G_2(z_2) G_1(z_1) \approx \frac{M_2(z_2) M_1(z_1)}{1 - \langle M_1(z_1) M_2(z_2)\rangle} \,. 
\end{equation}
To control \eqref{eq:m12}, we hence need to estimate the denominator of \eqref{eq:m12}, which is well known in case of $H_1 = H_2$, i.e.~$D_1 = D_2$ \cite{ETHpaper, equipart, pretherm}. Here, however, the analysis of \eqref{eq:m12} is much more intricate, since for general $D_1, D_2$ the deterministic approximations $M_1(z_1), M_2(z_2)$ do \emph{not} commute. In our main Proposition \ref{prop:stab}, we optimally track the dependence of \eqref{eq:m12} on the difference $D_1 - D_2$ of the two deformations and on $z_1-z_2$.

\subsection*{Notations}
For positive quantities $f,g$ we write $f\lesssim g$ (or $f=\mathcal{O}(g)$) and $f\sim g$ if $f \le C g$ or $c g\le f\le Cg$, respectively, for some constants $c,C>0$ which only depend on the constants appearing in the moment condition (see Assumption \ref{ass:Wigner}), the bound on $M$ in Assumption \ref{ass:M_bound}, or the constants from Assumption \ref{ass:H0}.
In informal explanations, we frequently use the notation $f \ll g$, which indicates that $f$ is "much smaller" than $g$. Moreover, we shall also write $w \approx z$ to indicate the closeness of $w, z \in \C$ with a not precisely specified error. The support of a function $f$ is denoted by $\mathrm{supp}(f)$. 

For any natural number $n$ we set $[n]: =\{ 1, 2,\ldots ,n\}$. Matrix entries are indexed by lowercase Roman letters $a, b, c, ...$ from the beginning of the alphabet. We denote vectors by bold-faced lowercase Roman letters ${\bm x}, {\bm y}\in\C ^N$, or lower case Greek letters $\psi, \phi \in \C^N$, for some $N\in\N$. Vector and matrix norms, $\lVert {\bm x}\rVert$ and $\lVert A\rVert$, indicate the usual Euclidean norm and the corresponding induced matrix norm. For any $N\times N$ matrix $A$ we use the notation $\langle A\rangle:= N^{-1}\mathrm{Tr}  A$ for its normalized trace and denote the spectrum of $A$ by $\sigma(A)$. 
Moreover, for vectors ${\bm x}, {\bm y}\in\C^N$ we denote their scalar product by $\langle {\bm x},{\bm y}\rangle:= \sum_{i} \overline{x}_i y_i$.

Finally, we use the concept of ``with very high probability'' \emph{(w.v.h.p.)} meaning that for any fixed $C>0$, the probability of an $N$-dependent event is bigger than $1-N^{-C}$ for $N\ge N_0(C)$. We also introduce the notion of \emph{stochastic domination} (see e.g.~\cite{semicirclegeneral}): given two families of non-negative random variables
\[
X=\left(X^{(N)}(u) : N\in\N, u\in U^{(N)} \right) \quad \mathrm{and}\quad Y=\left(Y^{(N)}(u) : N\in\N, u\in U^{(N)} \right)
\] 
indexed by $N$ (and possibly some parameter $u$  in some parameter space $U^{(N)}$), 
we say that $X$ is stochastically dominated by $Y$, if for all $\xi, C>0$ we have 
\begin{equation}
	\label{stochdom}
	\sup_{u\in U^{(N)}} \mathbf{P}\left[X^{(N)}(u)>N^\xi  Y^{(N)}(u)\right]\le N^{-C}
\end{equation}
for large enough $N\ge N_0(\xi,C)$. In this case we use the notation $X\prec Y$ or $X= \mathcal{O}_\prec(Y)$.

\subsection*{Acknowledgment} We thank Giorgio Cipolloni for helpful discussions in a closely related joint project. 

\section{Main results}
\label{sec:mainres}
The key players of our paper are \emph{deformed Wigner matrices}, i.e.~ matrices of the form $H = D + W$, where $D = D^* \in \C^{N \times N}$ is a bounded deterministic matrix, $\Vert D \Vert \le L$ for some $N$-independent $L> 0$ and $W = W^* \in \C^{N \times N}$ is a real symmetric or complex Hermitian Wigner matrices. This means, its entries are independently distributed random variables according to the laws\footnote{A careful examination of our proof reveals that the entries of $W$ need not be distributed identically. Indeed, only the matching of the second moments is necessary, but higher moments can differ.} $w_{ij} \stackrel{\mathrm{d}}{=} N^{-1/2}\chi_{\mathrm{od}}$ for $i < j$ and $w_{jj} \stackrel{\mathrm{d}}{=} N^{-1/2}\chi_{\mathrm{d}}$. For the single entry distributions we assume the following. 
		\begin{assumption}[Wigner matrix]  \label{ass:Wigner}
	We assume that $\chi_{\mathrm{d}}$ is a centered real random variable, and $\chi_{\mathrm{od}}$ is a real or complex random variable with $\E \chi_{\mathrm{od}} = 0$ and $\E |\chi_{\mathrm{od}}|^2 = 1$. Furthermore, we assume the existence of higher moments, namely	$\E |\chi_{\mathrm{d}}|^p + \E |\chi_{\mathrm{od}}|^p \le C_p$	for all $p\in \N$, where $C_p$ are positive constants.
\end{assumption}

It is well known \cite{slowcorr, firstcorr} that the resolvent of $H$, denoted by $G(z) := (H-z)^{-1}$ for $z \in \C\setminus \R$, becomes approximately deterministic in the large $N$ limit.  Its deterministic approximation (as a matrix) is given by $M(z)$, the unique solution of the Matrix Dyson equation (MDE)
\begin{equation} \label{eq:MDEfirst}
- \frac{1}{M(z)} = z - D + \langle M(z) \rangle \quad \text{for} \quad z \in \C\setminus \R \quad \text{under the constraint} \quad  \Im z \, \Im M(z) > 0 \,, 
\end{equation}
where $\Im M(z) := [M(z) - M(z)^*]/2 \ii$ and positivity is understood as a matrix. 
The corresponding  ($N$-dependent) \emph{self consistent density of states (scDos)} is defined as
\begin{equation}
	\label{eq:scdos}
	\rho(e):=\frac{1}{\pi}\lim_{\eta\downarrow 0}\langle \Im M(e+\ii\eta)\rangle\,.
\end{equation}
This is a compactly supported Hölder-$1/3$ continuous function on $\R$ which is in fact real-analytic on the 
set $\{ \rho > 0\}$\footnote{In~\cite{1506.05095,firstcorr,shape}, the scDos has been thoroughly analysed in increasing generality.
	It is supported on finitely many finite intervals and, roughly speaking, there are  three different regimes for the behavior or $\rho$: In the \emph{bulk}, $\rho$ is strictly positive; at the
	\emph{edge}, $\rho$ vanishes like a square root at the edges of every supporting interval which are well separated; at the \emph{cusp}, where two intervals of support (almost) meet, $\rho$ behaves (almost) as a cubic root. Correspondingly, $\rho$ is locally real analytic, H\"older-$1/2$, or H\"older-$1/3$ continuous,
	respectively. Near the singularities, it has  an approximately universal shape (see \eqref{eq:l_edge}--\eqref{eq:intern_min} in the proof of Lemma \ref{lem:int_stab}). \label{cusp}}. 
The positive harmonic extension of $\rho$ is denoted by $\rho(z) := \pi^{-1} |\langle \Im M(z) \rangle|$ for $z \in \C\setminus \R$.	We point out that not only the tracial quantity $\langle \Im M(e+\ii\eta)\rangle$ has an extension to the real axis, but the whole matrix $M(e) := \lim_{\eta\downarrow 0} M(e + \ii \eta)$ is well defined (see Lemma B.1~(b) of the \href{https://arxiv.org/abs/2301.03549}{arXiv: 2301.03549} version of~\cite{iid}). Moreover, for any small $\kappa > 0$ (independent of $N$) we define the \emph{$\kappa$-bulk} of the scDos \eqref{eq:scdos} as 
\begin{equation} \label{eq:bulk}
	\mathbf{B}_\kappa(\rho) = \left\{ x \in \R \; : \; \rho(x) \ge \kappa \right\}\,.
\end{equation}
It is a finite union of disjoint compact intervals, cf.~Lemma B.2 in
the \href{https://arxiv.org/abs/2301.03549}{arXiv: 2301.03549} version of~\cite{iid}. 
Note that, for $\Re z \in \mathbf{B}_\kappa$ it holds that $\Vert M(z) \Vert \lesssim 1$, as easily follows by taking the imaginary part of \eqref{eq:MDEfirst}.

Now, the resolvent $G$ is close to $M$ from \eqref{eq:MDEfirst} 
in the following \emph{averaged} and \emph{isotropic} sense:
\begin{equation}
	\label{eq:singlegllaw}
	|\langle (G(z)-M(z))B\rangle|\prec \frac{1}{N|\Im z|}, \qquad |\langle{\bm x}\,, (G(z)-M(z)) {\bm y}\rangle|\prec \frac{1}{\sqrt{N|\Im z|}} \,,
\end{equation}
uniformly in deterministic vectors $\lVert {\bm x}\rVert+\lVert{\bm y}\rVert\lesssim 1$ and deterministic matrices 
$\lVert B\rVert\lesssim 1$.  These estimates are called \emph{local laws} when $|\Im z|\ll 1$ and
\emph{global laws} when $|\Im z|\gtrsim 1$. 
To be precise about their validity, we recall that while
\eqref{eq:singlegllaw} hold for 
$\Re z \in \mathbf{B}_\kappa$ and $\mathrm{dist}(\Re z, \mathrm{supp}(\rho)) \gtrsim 1$ for \emph{arbitrary} 
bounded self-adjoint deformations $D =D^*$ (see \cite[Theorem~2.1]{slowcorr}), the complementary regime requires the additional Assumption \ref{ass:M_bound} on $D$ (see \cite[Theorem~2.6]{edgelocallaw} and \cite[Theorem~2.8]{cusplocallaw}). A sufficient condition for Assumption~\ref{ass:M_bound} is discussed in Remark \ref{rmk:Mbdd}; see also \cite{shape}.

In the remainder of this section, we formulate our main results on the two different Loschmidt echo scenarios described in Section \ref{sec:introduction}.

\subsection{Scenario I: Two deformations of a Wigner matrix} \label{subsec:twodef}
For the first echo scenario, we consider two deformed Wigner matrices, $H_j = D_j + W$, $j \in [2]$, and denote their resolvents and corresponding deterministic approximation \eqref{eq:MDEfirst} by $G_j$ and $M_j$, respectively.  
A natural definition of the averaged Loschmidt echo is 
\begin{equation} \label{eq:locechodef}
	\Losa(t) = \Losa_{H_1, H_2}^{(E_0, \eta_0)}(t):= \left| \frac{\left\langle \mathrm{e}^{\ii tH_1}\Im G_1(E_0 + \ii \eta_0) \mathrm{e}^{-\ii tH_2}\right\rangle}{\left\langle \Im M_1(E_0+ \ii \eta_0) \right\rangle} \right|^2\,,
\end{equation}
since  $\Im G/\langle \Im M \rangle$ in \eqref{eq:locechodef} effectively localizes around $E_0$ and \emph{averages} in a window of size $\eta_0 >0$. In Remark \ref{rmk:averaging} below we comment on the averaging implemented by \eqref{eq:locechodef}. Note that in order to match \eqref{eq:overintro}-\eqref{eq:echo1intro} from the introduction we need to replace $t$ by $-t$ in \eqref{eq:locechodef}. However, this replacement does not change the quantity \eqref{eq:locechodef} since
\begin{equation*}
\left\vert \left\langle \mathrm{e}^{\ii tH_1}\Im G_1(E_0 + \ii \eta_0) \mathrm{e}^{-\ii tH_2}\right\rangle\right\vert = \left\vert \left\langle \mathrm{e}^{\ii tH_2}\Im G_1(E_0 + \ii \eta_0) \mathrm{e}^{-\ii tH_1}\right\rangle\right\vert=\left\vert \left\langle \mathrm{e}^{-\ii tH_1}\Im G_1(E_0 + \ii \eta_0) \mathrm{e}^{\ii tH_2}\right\rangle\right\vert,
\end{equation*}  
where in the last step we used that $\ee^{\ii t H_1}$ and $\Im G_1(E_0+\ii\eta_0)$ commute. Using this observation we will work with \eqref{eq:locechodef} in the rest of the paper. The same comment applies also to the other versions of the
averaged Loschmidt echo defined in Section \ref{subsec:twodef}, namely to \eqref{eq:Av_Pr1}, \eqref{eq:Av_Pr2} and \eqref{eq:scavLos}.

We will henceforth assume that the deformations $D_1, D_2$ are such that the corresponding solutions $M_1, M_2$ to \eqref{eq:MDEfirst} are bounded. 
\begin{assumption}[Boundedness of $M$]\label{ass:M_bound} Let $D$ be an $N\times N$ Hermitian matrix and $M$ the solution to \eqref{eq:MDEfirst}. We assume that there exists an $N$-independent positive constant $L$ such that $\sup_{z\in\C \setminus \R} \lVert M(z)\rVert < L$. 
\end{assumption}

Assumption \ref{ass:M_bound} is the basis for the \emph{shape theory} of the scDos, which we briefly described in Footnote~\ref{cusp}. We now describe a sufficient condition on $D$ for Assumption \ref{ass:M_bound} to hold. 
 It basically requires that its ordered eigenvalue sequence has to be piecewise Hölder-$1/2$ continuous
as a function of the label.

\begin{remark}[Sufficient condition for Assumption \ref{ass:M_bound}] \label{rmk:Mbdd}
Denote the eigenvalues of any self-adjoint deformation $D$ by $\lbrace d_j\rbrace_{j=1}^N$ labeled in increasing order, $d_j\le d_k$ for $j<k$. Fix a (large) positive constant $L>0$. The set $\mathcal{M}_L$ of admissible self-adjoint deformations $D$ is defined as follows: we say that $D\in \mathcal{M}_L$ if $\lVert D\rVert\le L$ and there exists an $N$-independent partition $\lbrace I_s\rbrace_{s=1}^m$ of $[0,1]$ in at most $L$ segments such that for any $s\in[1,m]$ and any $j,k\in[1,N]$ with $j/N,k/N\in I_s$ we have $\vert d_j-d_k\vert \le L\vert j/N- k/N\vert^{1/2}$. Since the operator $\mathcal{S}=\langle\cdot\rangle$ is flat, condition $D\in\mathcal{M}_L$ implies that $D$ satisfies Assumption \ref{ass:M_bound} for some $L'<\infty$ by means of \cite[Lemma~9.3]{shape}.
\end{remark}

We can now formulate our first main result.

\begin{theorem}[Averaged Loschmidt echo with two deformations] \label{thm:echo1} Let $W$ be a Wigner matrix satisfying Assumption \ref{ass:Wigner}, and $D_1, D_2\in \C^{N \times N}$ be bounded, traceless\footnote{If $D_1$ or $D_2$ had a non-zero trace, it could be absorbed by a simple (scalar) energy shift.} Hermitian matrices, i.e.~$\lVert D_j\rVert\le L$ for some $L > 0$ and $\langle D_1 \rangle = \langle D_2 \rangle = 0$, additionally satisfying Assumption \ref{ass:M_bound}. 
	Let $E_0$ be an energy in the bulk of the scDos of $H_1$ and $H_2$, i.e.~assume that there exist $\delta, \kappa>0$ such that $[E_0-\delta,E_0+\delta] \subset  \mathbf{B}_\kappa(\rho_1) \cap \mathbf{B}_\kappa(\rho_2)$. 

Consider the deformed Wigner matrices $H_j := D_j + W$ for $j \in [2]$ and the corresponding averaged (at energy $E_0$ in a window of size $\eta_0> 0$) Loschmidt echo $\Losa(t)$ for times $t \ge 0$ defined in \eqref{eq:locechodef}. 
Then we have the following: 
	\begin{itemize}
	\item[(i)] \textnormal{[Short-time parabolic decay]} As $t \to 0$, it holds that 
	\begin{equation} \label{eq:echoshort}
		\Losa(t) = 1 -  \gamma  t^2+ \mathcal{O}\big(\langle D^2 \rangle t^3\big) + \mathcal{O}_\prec((N \eta_0)^{-1}) 
	\end{equation}
	where the \emph{decay parameter} is given by 
	 $\gamma := \langle \left(D - \langle P D \rangle \right)^2 P\rangle$, where we abbreviated $D:= D_2 - D_1$ and $ P := \Im  M_1(E_0 + \ii \eta_0)/\langle \Im M_1 (E_0 + \ii \eta_0) \rangle$. It satisfies $\gamma \sim \Delta^2 := \langle D^2 \rangle$ and the implicit constant depends only on $\kappa$ and $L$. 
 
	The implicit  constants in the error terms in \eqref{eq:echoshort} depend only on $L, \delta, \kappa$ and the $C_p$'s from Assumption~\ref{ass:Wigner}. 
	\item[(ii)] \textnormal{[Intermediate-time asymptotic decay]} 
	Take a (large) positive $K$ and consider times $1\le t\le K/\Delta^2$. Then there exists a positive constant $c$ such that whenever $\Delta <c$ and $\eta_0<\Delta/\vert\log \Delta\vert$ it holds that
	\begin{equation}
		\Losa(t) = \mathrm{e}^{- \Gamma t} + \mathcal{O}\left(\mathcal{E}\right) + \mathcal{O}_\prec\big(C(t)/N\big),
		\label{eq:echolong}
	\end{equation}
	where the \emph{rate} $\Gamma$ (explicitly given in \eqref{eq:Gamma}) satisfies $\Gamma \sim \Delta^2$ with the implicit constant depending only on $\kappa$ and $L$. Moreover, we denoted 
	\begin{equation}
		\mathcal{E} = \mathcal{E}(t, \Delta, \eta_0) = \frac{1+\log t}{t}+\Delta\vert\log \Delta\vert+\frac{\eta_0\vert\log\Delta\vert}{\Delta}
	\label{eq:err_thm1}
	\end{equation}
	and $C(t) > 0$ is a positive constant depending only on $t$. 
	
The implicit constants in the error terms in \eqref{eq:echolong} depend only on $L,\delta,\kappa,K$ and the  $C_p$'s from Assumption~\ref{ass:Wigner}. 
\end{itemize}
\end{theorem}
Since $t \le K/\Delta^2$, we find that the leading term in \eqref{eq:echolong} remains of order one throughout the whole time regime. The error term $\mathcal{E}$ is small compared to this leading term if $t \gg 1$, $\Delta \ll 1$, and $\eta_0 \ll \Delta/|\log \Delta|$, hence these relations define the regime of the parameters where our theorem is meaningful.

The following corollary to Theorem \ref{thm:echo1} reveals the key property of the Loschmidt echo, the partial recovery of the initial overlap, as discussed in the introduction; see Figure \ref{fig:2}. 
\begin{corollary}[Averaged Loschmidt echo process]\label{cor:2bumps} Assume the conditions of Theorem \ref{thm:echo1}. For time $t>0$ define the \emph{averaged Loschmidt echo process} $\Epa_t(s)$, $s\in [0,2t]$, as follows:  
\begin{subequations}
\begin{alignat}{2}
&\label{eq:Av_Pr1}\Epa_t(s):=\left\vert \frac{\left\langle \ee^{\ii sH_1}\Im G_1(E_0+\ii\eta_0)\right\rangle}{\left\langle\Im M_1(E_0+\ii\eta_0)\right\rangle}\right\vert^2,\quad&& s\in[0,t],\\
&\label{eq:Av_Pr2}\Epa_t(s):=\left\vert \frac{\left\langle \ee^{\ii tH_1}\Im G_1(E_0+\ii\eta_0)\ee^{-\ii(s-t)H_2}\right\rangle}{\left\langle\Im M_1(E_0+\ii\eta_0)\right\rangle}\right\vert^2,\quad&& s\in(t,2t].
\end{alignat}
\end{subequations}
Let $\lim^*$ be the simultaneous limit in $\Delta, \eta_0, t$ such that $\Delta, \eta_0\to 0$ and $t\to\infty$ under constraints $\Delta^2\ll \eta_0\ll \Delta/\vert \log \Delta\vert$ and $1/\eta_0\ll t\lesssim 1/\Delta^2$. Here $a\ll b$ means that $a/b\to 0$ in this limit. Then almost surely we have
\begin{equation}
\mathrm{lim}^*\limsup_{N\to\infty}\frac{\Epa_t(t)}{\Epa_t(2t)}=\mathrm{lim}^*\limsup_{N\to\infty}\frac{\Epa_t(t)}{e^{-\Gamma t}} = 0,
\label{eq:2bumps}
\end{equation}
where $\Gamma$ is the same as in Theorem \ref{thm:echo1}.
\end{corollary}

\begin{proof}[Proof of Corollary \ref{cor:2bumps}] Firstly take the limit $N\to\infty$ in the denominator $\Epa_t(2t)=\overline{\mathfrak{M}}(t)$ of \eqref{eq:2bumps}. Recall the definition of $\mathcal{E}$ from \eqref{eq:err_thm1}. By means of Theorem \ref{thm:echo1} we have
\begin{equation*}
\liminf_{N\to\infty}\Epa_t(2t) = \liminf_{N\to\infty}\left(\ee^{-\Gamma t}+\mathcal{O}\left(\mathcal{E}(t,\Delta,\eta_0)\right)\right)=\liminf_{N\to\infty}\left(\ee^{-\Gamma t}(1+o(1))\right)\sim 1
\end{equation*}
in the limit $\lim^*$. Here we used that $\Gamma\sim \Delta^2$ and $t\lesssim \Delta^{-2}$, so $\ee^{-\Gamma t}\sim 1$. Thus in order to verify \eqref{eq:2bumps} it is sufficient to show that
\begin{equation*}
\mathrm{lim}^*\limsup_{N\to\infty}\Epa_t(t)=0.
\end{equation*}
From the average single resolvent global law for $H_1$, see \eqref{eq:singlegllaw} or
 \cite[Theorem~2.1]{slowcorr}, we get that
\begin{equation*}
\lim_{N\to\infty} \left\vert \left\langle \ee^{\ii tH_1}\Im G_1(E_0+\ii\eta_0)\right\rangle - \int_\R \ee^{\ii tx}\frac{\eta_0}{(x-E_0)^2+\eta_0^2}\rho_1(x)\dif x\right\vert = 0.
\end{equation*}
Recall that $E_0\in{\bm B}_\kappa(\rho_1)$. Thus $\langle \Im M(E_0+\ii\eta_0)\rangle\sim 1$ for $\eta_0\to 0$ and
\begin{equation}
\limsup_{N\to\infty}\Epa_t(t)\lesssim\limsup_{N\to\infty} \left\vert \int_\R \ee^{\ii tx}\frac{\eta_0}{(x-E_0)^2+\eta_0^2}\rho_1(x)\dif x\right\vert^2 \lesssim \left(\frac{1}{\eta_0 t}\right)^2.
\label{eq:eta_0-smooth}
\end{equation}
In the last inequality we employed integration by parts. Additionally we used that $\rho_1(x)$ is a bounded function of $x$ which is guaranteed by Assumption \ref{ass:M_bound} and that $\rho_1(x)$ has bounded derivative for $|x-E_0| \le \delta$ (see also Footnote \ref{cusp}), where $\delta$ was fixed in Theorem \ref{thm:echo1}. Both of this bounds (on $\rho_1(x)$ and $\dif\rho_1(x)/\dif x$) are uniform in $N$. In the limit $\lim^*$ we have $\eta_0 t\to \infty$, so \eqref{eq:eta_0-smooth} finishes the proof of Corollary~\ref{cor:2bumps}.
\end{proof}

As mentioned in the introduction, we also have the following corollary to Theorem \ref{thm:echo1}. 

\begin{corollary}[Scrambled averaged Loschmidt echo with two deformations] \label{cor:echo1}
Assume the conditions of Theorem \ref{thm:echo1} and consider (as a variant of \eqref{eq:locechodef}) the \emph{scrambled} averaged Loschmidt echo
\begin{equation} \label{eq:scavLos}
\Losa_\delta^{\rm sc}(t) :=  \left| \frac{\left\langle \ee^{- \ii \delta \widetilde{W}}\mathrm{e}^{\ii tH_1}\Im G_1(E_0 + \ii \eta_0) \mathrm{e}^{-\ii tH_2}\right\rangle}{\left\langle \Im M_1(E_0+ \ii \eta_0) \right\rangle} \right|^2\,,
\end{equation}
where $\widetilde{W}$ is a Wigner matrix satisfying Assumption \ref{ass:Wigner}, \emph{independent} of $W$. Moreover, let $\varphi$ be the Fourier transform of the semi-circular density of states $\rho_{\mathrm{sc}}(x) := (2 \pi)^{-1} \sqrt{[4-x^2]_+}$, which is explicitly given as
	\begin{equation} \label{eq:schat}
			\varphi(\delta) := \widehat{\rho_{\mathrm{sc}}}(\delta) = \int_\R \mathrm{e}^{- \ii \delta x} \rho_{\mathrm{sc}}(x) \rd x = \frac{J_1(2 \delta)}{\delta}
		\end{equation}
	where $J_1$ is the first order Bessel function of the first kind. 
	
	Then, instead of \eqref{eq:echoshort}--\eqref{eq:echolong}, we have that 
	\begin{equation*}
	\Losa_\delta^{\rm sc}(t) =
	(\varphi(\delta))^2 \, \big[1 - \gamma t^2 + \mathcal{O}\big(\langle D^2 \rangle t^3\big) + \mathcal{O}_\prec((N \eta_0)^{-1}) \big] + \mathcal{O}_\prec\big(\delta/(N \eta_0)\big) \quad \text{as} \quad t \to 0
	\end{equation*}
	and
		\begin{equation*}
		\Losa_\delta^{\rm sc}(t) = 
			(\varphi(\delta))^2 \,	\big[ \mathrm{e}^{- \Gamma t} + \mathcal{O}\left(\mathcal{E}\right) + \mathcal{O}_\prec\big(C(t)/N\big)\big]  + \mathcal{O}_\prec\big(\delta/(N \eta_0)\big) \quad \text{for} \quad 1 \le t \le K/\Delta^2
	\end{equation*}
in the short and intermediate time regimes, respectively. 
\end{corollary}

\begin{proof}[Proof of Corollary \ref{cor:echo1}] Denote $A := \mathrm{e}^{\ii tH_1}\Im G_1(E_0 + \ii \eta_0) \mathrm{e}^{-\ii tH_2}$ and observe that $\Vert A \Vert \le 1/\eta_0$. Then, by residue calculus with the contour $C_\delta := \{ z \in \C : \mathrm{dist}(z, [-2,2]) = \delta^{-1}\}$ and a single resolvent law\footnote{To be precise, when $\delta < 1$, we use the slightly improved average \emph{global} law $|\langle A(W-z)^{-1}  \rangle - m(z) \langle A \rangle| \prec \delta^2 \Vert A \Vert/N$ (see, e.g., \cite[Theorem~2.1]{slowcorr}).} as in \eqref{eq:singlegllaw}, using only the randomness of $\widetilde{W}$, we find  
\begin{equation*}
	\begin{split}
		\langle \mathrm{e}^{- \ii \delta \widetilde{W}} A \rangle &= \frac{1}{2 \pi \ii} \oint_{C_\delta} \mathrm{e}^{-\ii \delta z } \langle A (W-z)^{-1} \rangle \rd z = \frac{\langle A \rangle}{2 \pi \ii} \oint_{C_\delta} \mathrm{e}^{-\ii \delta z } m_{\mathrm{sc}}(z) \rd z + \mathcal{O}_\prec(\delta/(N \eta_0))  \\
		&= \langle A \rangle\int_\R \mathrm{e}^{- \ii \delta x} \rho_{\mathrm{sc}}(x) \rd x + \mathcal{O}_\prec(\delta/\sqrt{N}) = \langle A \rangle \varphi(\delta)  + \mathcal{O}_\prec(\delta/(N \eta_0)) \,. 
	\end{split}
\end{equation*}
The rest of the proof follows from Theorem \ref{thm:echo1}.
\end{proof}

We close this section by commenting on the small averaging of the Loschmidt echo
over several energy states implemented in \eqref{eq:locechodef}. This is a necessary technical step
for our proof in Scenario I that relies on a two-resolvent local law. Note that averaging will not be 
necessary for Scenario II since it uses only single resolvent local law.

\begin{remark}[Averaging of the Loschmidt echo] \label{rmk:averaging} We provide two independent non-rigorous arguments for the averaged Loschmidt echo $\Losa$ and the non-averaged Loschmidt echo $\Los$ being close to each other. 
	
	\begin{itemize}
\item[(1)] 	First, by means of the Eigenstate Thermalization Hypothesis (ETH) for a deformed Wigner matrix $H = D + W$, see \cite[Theorem~2.7]{equipart}, and a single resolvent local law \eqref{eq:singlegllaw}, it holds that 
\begin{equation} \label{eq:ETHarg}
	\langle \bm u_j, A \bm u_j \rangle \approx \frac{\langle \Im M(E_0 + \ii \eta_0) A \rangle}{\langle \Im M(E_0 + \ii \eta_0)\rangle} \approx \frac{\langle \Im G(E_0 + \ii \eta_0) A \rangle}{\langle \Im M(E_0 + \ii \eta_0)\rangle}\,. 
\end{equation}
Here, $A$ is an arbitrary deterministic matrix, $\bm u_j$ is a (normalized) eigenvector of $H$ with eigenvalue $\approx E_0$, and $\eta_0$ a small regularization. In this sense, the pure state $\ket{\bm u_j} \bra{\bm u_j}$ is \emph{weakly} close to $\Im G/\langle \Im M \rangle$ (i.e.~if tested against a deterministic $A$), which 
heuristically supports the implementation of the averaged Loschmidt echo in \eqref{eq:locechodef}. 
However, the rigorous ETH statements do not allow to choose  $A$  depending on the underlying randomness
like $A=e^{-\ii t H_2} e^{\ii t H_1}$. 
\item[(2)] 	Another supporting argument uses the fact that
the \emph{averaged overlap function}  
$\Overa^{(E, \eta_0)}(t)$ (in particular its phase)
is approximately constant as long as $|E-E_0| \lesssim \eta_0$. Hence it is irrelevant if one (a) first averages and then takes absolute value square, or (b) does it the other way around. We claim that $\Overa^{(E, \eta_0)}(t)$ is approximately constant as $E$ varies. This follows by a simple computation using that (i) $\Overa^{(E_0, \eta_0)}(t) \approx I_{E_0, \eta_0}(t)/\langle \Im M_1(E_0 + \ii \eta_0) \rangle$ (see \eqref{eq:rescalcstart} and \eqref{eq:detcontint}), (ii) $I_{E_0,\eta_0}$ is given by $\ee^{\ii t\ms_0} \langle \Im M_1(E_0 + \ii \eta_0) \rangle$ (see \eqref{Iphase}), (iii) the exponent $\ms_0$ is Lipschitz continuous on scale $\Delta$ (see the last relation of \eqref{eq:deriv_bounds}), and (iv) we have $t \lesssim \Delta^{-2}$ and $\eta_0\ll\Delta$ by assumption.
	\end{itemize}

	Both, the ETH argument \eqref{eq:ETHarg} and  the fact that $\Overa^{(E, \eta_0)}(t)$ is approximately constant as long as $|E-E_0| \lesssim \eta_0$, independently indicate that the averaged Loschmidt echo $\Losa$ and the non-averaged Loschmidt echo $\Los$ should practically agree with each other. However, neither of them constitutes a rigorous proof, since (1) the observable $A$ in \eqref{eq:ETHarg} cannot be chosen to depend on the randomness, and (2) we cannot exclude 
that for some initial fixed energy state $\psi_0$, $\Over$ in  \eqref{eq:overintro} behaves very differently from its typical 
value computed by local averaging. 
\end{remark}

\subsection{Scenario II: Perturbation by a Wigner matrix} \label{subsec:onedef}

For the second echo scenario, we consider a single deformed Wigner matrix $H_\lambda = H_0 + \lambda W$ and the Loschmidt echo
\begin{equation} \label{eq:echoprot2}
	\Los(t) = \Los_{H_\lambda, H_0}^{(E_0, \Delta)}(t) := \left| \langle \psi_0, \mathrm{e}^{\ii t H_\lambda} \mathrm{e}^{- \ii t H_0} \psi_0 \rangle \right|^2
\end{equation}
for some normalized initial state $\psi_0 \in \C^N$ with energy $E_0 = \langle \psi_0, H_0 \psi_0 \rangle$ and localized in an interval of size $\Delta$ around $E_0$ (see Assumption \ref{ass:state} below for a precise statement). The localization parameter $\Delta$ plays the same role as $\eta_0$ in Section \ref{subsec:twodef}, but here we work with a sharp cutoff in the energy.

The unperturbed Hamiltonian $H_0$ is assumed to satisfy the following. 
\begin{assumption}[$H_0$ and its limiting density of states] \label{ass:H0}
	The Hamiltonian $H_0 $ is deterministic, self-adjoint $H_0 = H_0^*$, and uniformly bounded, $\Vert H_0 \Vert \le C_{H_0}$ for some $C_{H_0} > 0$. We denote the resolvent of $H_0$ at any spectral parameter $z \in \C\setminus \R$ by $M_0(z):= (H_0 -z)^{-1}$.
	Moreover, we assume the following: 
	\begin{itemize}
		\item[(i)] 	There exists a compactly supported measurable function $\rho_0 : \R \to [0,+\infty)$ with 
		$\int_\R \rho_0(x) \rd x = 1$
		and two positive sequences $\epsilon_0(N)$ and $\eta_0(N)$, both converging to zero as $N\to\infty$,
		such that, uniformly in $z \in \C\backslash\R$ with $\eta:=|\Im z| \ge \eta_0 \equiv \eta_0(N)$, we have
		\begin{equation} \label{eq:rho0}
			\langle M_0(z)\rangle = m_0(z) + \mathcal{O}(\epsilon_0) \quad \text{with} \quad \epsilon_0 \equiv \epsilon_0(N)\,.
		\end{equation}
		Here, 
		\begin{equation} \label{eq:m0}
			m_0(z):= \int_\R \frac{\rho_0(x)}{x-z}\mathrm{d}x
		\end{equation} 
		is the Stieltjes transform of $\rho_0$. We refer to $\rho_0$ as the \emph{limiting density of states}, and to $\mathrm{supp}(\rho_0)$ as the \emph{limiting spectrum} of $H_0$. 
		\item[(ii)] For small positive constants $\kappa,c>0$, we define the set of \emph{admissible energies} $\sigma_{\mathrm{adm}}^{(\kappa,c)}$ in the limiting spectrum of $H_0$ by\footnote{Here, $C^{1,1}(J)$ denotes the set of continuously differentiable functions with a Lipschitz-continuous derivative on an interval $J$, equipped with the norm $\lVert f\rVert_{C^{1,1}(J)} := \lVert f\rVert_{C^1(J)} + \sup\limits_{x,y\in J: x\neq y} \frac{|f'(x)-f'(y)|}{|x-y|}$.}
		\begin{equation} \label{eq:admiss_spec}
			\sigma_{\mathrm{adm}}^{(\kappa,c)} := \left\{x \in \mathrm{supp}(\rho_0) : \inf_{|y-x|\le\kappa}\rho_0(y) > c,\, \lVert\rho_0\rVert_{C^{1,1}([x-\kappa, x+\kappa])} \le 1/c\right\}.
		\end{equation}
		We assume that for some positive $\kappa,c> 0$, $\sigma_{\mathrm{adm}}^{(\kappa,c)}$ is not empty.
	\end{itemize}
\end{assumption}
Assuming that the set of admissible energies in \eqref{eq:admiss_spec} is non-empty guarantees the limiting spectrum $\mathrm{supp} \rho_0$ has a part, where the limiting density of states behaves regularly, i.e. it is sufficiently smooth and strictly positive (in the \emph{bulk}). 

\begin{assumption}[Locality of the initial state] \label{ass:state}
	Given Assumption \ref{ass:H0}, we first pick a \emph{reference energy}
	\begin{equation} \label{eq:E0}
		E_0 \in \sigma_{\mathrm{adm}}^{(\kappa_0,c_0)} \quad  \text{for some}  \quad \kappa_0, c_0>0,
	\end{equation}
	and further introduce $I_\delta:=[E_0-\delta,E_0+\delta]$  for any $0<\delta<\kappa_0$. 
	Moreover, take an \emph{energy width} $\Delta \in (0, \kappa_0/2)$ and let
	$\Pi_\Delta := \mathbf{1}_{I_\Delta}(H_0)$ be the spectral projection of 
	$H_0$ onto the interval $I_\Delta$. 
	Then, we assume that the initial state $\psi_0 \in \C^{N}$ is normalized, $\Vert \psi_0 \Vert = 1$, has energy $E_0 = \langle \psi_0, H_0 \psi_0 \rangle$, and satisfies $\Pi_\Delta \psi_0 = \psi_0$, i.e.~$\psi_0$ is localized in~$I_\Delta$. 
\end{assumption}

\begin{theorem}[Loschmidt echo with a single deformation] \label{thm:echo2}
	Consider the Loschmidt echo \eqref{eq:echoprot2} for times $t \ge 0$ and assume that its constituents satisfy Assumptions \ref{ass:Wigner} and \ref{ass:H0}--\ref{ass:state}. Then we have the following: 
	\begin{itemize}
		\item[(i)] \textnormal{[Short-time parabolic decay]} As $t \to 0$ it holds that 
		\begin{equation} \label{eq:echoshort2}
			\Los(t) = 1 - \lambda^2 t^2 + \mathcal{O}(\lambda^2 t^3) + \mathcal{O}_\prec\big( 1/\sqrt{N}\big) \,. 
		\end{equation}
		
		The implicit constants in the error terms in \eqref{eq:echoshort2} only depend on $C_{H_0}$ and the $C_p$'s from Assumption~\ref{ass:Wigner}. 
		\item[(ii)] \textnormal{[Intermediate-time asymptotic decay]} For all times $t \ge 0$ it holds that 
		\begin{equation} \label{eq:echolong2}
			\Los(t) = \mathrm{e}^{-2\pi \rho_0(E_0) \lambda^2 t} + \mathcal{O}(\mathcal{E}) + \mathcal{O}_\prec\big(C(t, \lambda)/\sqrt{N}\big) \,,
		\end{equation}
		where for any fixed $T > 0$ the error term $\mathcal{E}$, explicitly given in \eqref{eq:Ereg}, satisfies
		\begin{equation*}
			\lim\limits_{\Delta \to 0} \lim\limits_{\substack{t \to \infty, \lambda \to 0 \\ \lambda^2 t \le T}} \lim\limits_{N \to \infty} \mathcal{E} = 0 
		\end{equation*}
		and the constant $C(t, \lambda)> 0$ depends only on its arguments. 
		
		The implicit constants in the error terms in \eqref{eq:echolong2} depend only on $C_{H_0}$ from Assumption \ref{ass:H0}, $\kappa_0, c_0$ from Assumption \ref{ass:state}, and the $C_p$'s from Assumption \ref{ass:Wigner}. 
	\end{itemize}
\end{theorem}
In the small time regime, $t \to 0$, \eqref{eq:echoshort2} is surely more precise than \eqref{eq:echolong2}, which, in turn, describes exponential decay for times of order $t \sim \lambda^{-2}$. 

The proof of the following Corollary \ref{cor:echo2} is completely analogous to the proof of Corollary \ref{cor:echo1} (only using an isotropic law instead of an averaged law) and so omitted. 
\begin{corollary}[Scrambled Loschmidt echo with a single deformation] \label{cor:echo2}
Assume the conditions of Theorem~\ref{thm:echo2} and consider (as a variant of \eqref{eq:echoprot2}) the \emph{scrambled} Loschmidt echo
\begin{equation} \label{eq:scLos}
	\Los_\delta^{\rm sc}(t) :=  \left| \left\langle \psi_0, \mathrm{e}^{\ii t H_\lambda} \ee^{-\ii \delta \widetilde{W}}\mathrm{e}^{- \ii t H_0} \psi_0 \right\rangle \right|^2\,, 
\end{equation}
where $\widetilde{W}$ is a Wigner matrix satisfying Assumption \ref{ass:Wigner}, \emph{independent} of $W$. Moreover, let $\varphi$ be given by~\eqref{eq:schat}. 

Then, instead of \eqref{eq:echoshort2}--\eqref{eq:echolong2}, we have that 
\begin{equation*}
	\Los_\delta^{\rm sc}(t) =
	(\varphi(\delta))^2 \, \big[1 - \lambda^2 t^2 + \mathcal{O}(\lambda^2 t^3) + \mathcal{O}_\prec\big( 1/\sqrt{N}\big) \big] + \mathcal{O}_\prec\big(\delta/\sqrt{N}\big) \quad \text{as} \quad t \to 0
\end{equation*}
and
\begin{equation*}
	\Los_\delta^{\rm sc}(t) = 
	(\varphi(\delta))^2 \,	\big[ \mathrm{e}^{-2\pi \rho_0(E_0) \lambda^2 t} + \mathcal{O}(\mathcal{E}) + \mathcal{O}_\prec\big(C(t, \lambda)/\sqrt{N}\big)\big]  + \mathcal{O}_\prec\big(\delta/\sqrt{N}\big) \quad \text{for} \quad \lambda^2 t \le T
\end{equation*}
respectively. 
\end{corollary}

The rest of the paper is devoted to proving Theorems \ref{thm:echo1} and \ref{thm:echo2}. The proof of Theorem \ref{thm:echo1} is conducted in Sections \ref{sec:pfecho1short}--\ref{sec:pfecho1long}. In Section \ref{sec:pfecho23} we prove Theorem \ref{thm:echo2}. The proof of several technical results from Section \ref{sec:pfecho1long} is deferred to Sections \ref{sec:stabopshiftpf} and \ref{sec:conintproof}, and Appendix \ref{app:technical}. 

\section{Short-time parabolic decay in Scenario I: Proof of Theorem \ref{thm:echo1}~(i)} \label{sec:pfecho1short}
In the following, we abbreviate $\widetilde{P} = \Im G_1(E_0 + \ii \eta_0)/\langle \Im M_1(E_0 + \ii \eta_0) \rangle$, such that  $\Los(t)$ can be written as
\begin{equation} \label{eq:specdecshort}
	\Losa(t) = \left| \left\langle \mathrm{e}^{\ii tH_1} \widetilde{P} \mathrm{e}^{-\ii tH_2}\right\rangle \right|^2 = \left| \left\langle \widetilde{P}\mathrm{e}^{\ii tH_1} \mathrm{e}^{-\ii tH_2}\right\rangle \right|^2 \,. 
\end{equation}

Next, we trivially Taylor expand $\mathrm{e}^{ \ii tH_1}$ and $ \mathrm{e}^{-\ii t H_2}$ to second order, leaving us with 
\begin{equation} \label{eq:preBCH}
	\mathrm{e}^{ \ii tH_1} \mathrm{e}^{-\ii t H_2} = 1 + \ii t(H_1 - H_2) - \frac{t^2}{2} \big( (H_1 -H_2)^2 - [H_1, H_2] \big) + \mathcal{O}(t^3) \,. 
\end{equation}
Plugging this in \eqref{eq:specdecshort}, we find 
\begin{equation} \label{eq:parabolic}
	\begin{split}
		\Losa(t) = &\left\langle \widetilde{P}  \left(1 - \tfrac{t^2}{2}((H_1 - H_2)^2 - [H_1, H_2])\right) \right\rangle^2 \\
		& \qquad + t^2 \langle \widetilde{P} (H_1 -H_2) \rangle^2 + \mathcal{O}(t^3) + \mathcal{O}_\prec((N \eta_0)^{-1}) \\ 
		= &1 - \big\langle \left(D - \langle P D \rangle \right)^2 P \big\rangle \, t^2 + \mathcal{O}(t^3) + \mathcal{O}_\prec((N \eta_0)^{-1}) \,. 
	\end{split}
\end{equation}
Here, we additionally used that $\langle \widetilde{P} [H_1,H_2]  \rangle = 0$ since $\widetilde{P} $ is a function of $H_1$, $D= H_2 - H_1$, and a single resolvent law in the form $\langle \widetilde{P} A \rangle  = \langle P A \rangle  + \mathcal{O}_\prec((N \eta_0)^{-1})$ for any $A$ with $\Vert A \Vert \lesssim 1$. The fact that the decay parameter $\gamma = \langle \left(D - \langle P D \rangle \right)^2 P\rangle$ satisfies $\gamma \sim \Delta^2$ is a simple consequence of the \emph{flatness} of the stability operator for a deformed Wigner matrix (see, e.g., \cite[Proposition~3.5]{shape}). 

In order to conclude \eqref{eq:echoshort}, it remains to show that the error term $\mathcal{O}(t^3)$ in \eqref{eq:parabolic} is actually improvable to $\mathcal{O}(\langle D^2\rangle t^3)$. To see this, we (formally)\footnote{In order to guarantee convergence of the BCH expansion \eqref{eq:BCH}, we need the time $t$ to be small enough such that $|t|(\Vert H_1 \Vert + \Vert H_2 \Vert) < \log 2$ \cite{Suzuki1977}, which can be achieved in an open interval around zero, since $\Vert D_i\Vert \lesssim 1$ and $\Vert W \Vert \le 2 + \epsilon$ with very high probability.} employ the Baker-Campbell-Hausdorff (BCH) formula, to write the exponentials as 
\begin{equation} \label{eq:BCH}
	\begin{split}
		\mathrm{e}^{ \ii tH_1} \mathrm{e}^{-\ii t H_2} =& \,  \mathrm{e}^{K} \quad \text{with} \\
		K =& \,  \ii t (H_1 - H_2) + \frac{t^2}{2} [H_1 , H_2] + \frac{\ii t^3}{12} \big( [H_1, [H_1, H_2]]  - [H_2, [H_1, H_2]]\big) \\
		&- \frac{t^4}{24} [H_2, [H_1, [H_1, H_2]]] + ... 
	\end{split}
\end{equation}
and note that every summand in the expression for $K$ in \eqref{eq:BCH} can be written as a linear combination of nested commutators of $D$ with $H \equiv H_1$ with one $D$ always being in the innermost commutator. Hence, to conclude the desired, we need to show that, (i) all the terms in $\mathrm{e}^K$ containing only a single $D$ vanish, when evaluated in $\langle \widetilde{P} ... \rangle$, and (ii) all the terms in $\mathrm{e}^K$ containing at least two $D$'s lead to an additional $\langle D^2\rangle$-factor in the error term.

For (i), note that the only way to have just a single $D$ in a nested commutator is precisely $\mathrm{ad}_H^n(D)$ with $\mathrm{ad}_H(D) := [H,D]$. Evaluated in $\langle \widetilde{P} ... \rangle$, this vanishes, $\langle \widetilde{P} \mathrm{ad}_H^n(D)  \rangle = 0$, since $[\widetilde{P}, H] = 0$ and hence 
\begin{equation} \label{eq:algcancpara}
	\langle \widetilde{P} \mathrm{ad}_H^n(D)  \rangle= \sum_{k=0}^n \binom{n}{k} (-1)^k \langle \widetilde{P}  H^{n-k} D H^k  \rangle = \langle \widetilde{P} H^n D  \rangle \sum_{k=0}^n \binom{n}{k} (-1)^k = 0\,. 
\end{equation}
For (ii), we take a product, say, $T$, of $H$'s and at least two $D$'s, resulting from resolving (a product of) nested commutators, and estimate
\begin{equation} \label{eq:D2higher}
	\begin{split}
		\left| \langle \widetilde{P} T \rangle \right| \lesssim \langle \widetilde{P} D^2 \rangle = \langle P D^2 \rangle + \mathcal{O}_\prec\big((N \eta_0)^{-1}\big) \lesssim \langle D^2 \rangle + \mathcal{O}_\prec\big((N \eta_0)^{-1}\big)\,. 
	\end{split}	
\end{equation}
In the first step, we estimated all $H$'s and all but two $D$'s in $T$ by their operator norm, additionally using that $\widetilde{P} \ge 0$ and $[H, \widetilde{P}] = 0$. In the second step, we employed the single resolvent law \eqref{eq:singlegllaw}, while in the last step we used $\Vert P\Vert \lesssim 1$. 

We have hence shown, that all the terms of $\mathrm{e}^K$ in \eqref{eq:BCH} carrying at least a third power of $t$, can in fact be bounded with an additional $\langle D^2 \rangle$-factor compared to \eqref{eq:parabolic}. This concludes the proof. \qed

\section{Asymptotic decay in Scenario I: Proof of Theorem \ref{thm:echo1}~(ii)} \label{sec:pfecho1long}

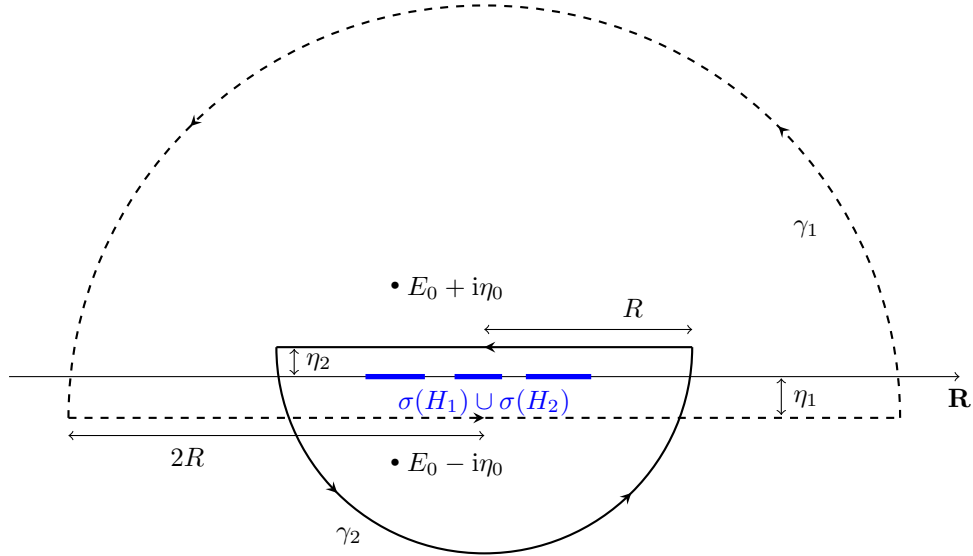
\begin{figure}[h]
	\begin{center}
		\begin{tikzpicture}[scale=\textwidth/19.2cm]
			\nc
			\draw[black,->] (-8,0) -- (8,0);
			\draw (8,0) node[below=1pt] {\nc$\R$};
			
			\draw[blue,line width=0.7mm] (-2,0) -- (-1,0);
			\draw[blue,line width=0.7mm] (-0.5,0) -- (0.3,0);
			\draw[blue,line width=0.7mm] (0.7,0) -- (1.8,0);
			\draw (0,0) node[below=1pt] {\color{blue}$\sigma(H_1) \cup \sigma(H_2)$\nc};
			
			\draw (5,2.5) node[right=1pt] {\nc $\gamma_1$};
			\draw (-2.7,-2.7) node[right=1pt] {\nc $\gamma_2$};
			
			\draw[black,thick,dashed,postaction={on each segment={mid arrow=black}}] (7,-0.7) arc[start angle=0, end angle=180, radius=7];
			\draw[black,thick,dashed,postaction={on each segment={mid arrow=black}}] (-7,-0.7) -- (7,-0.7);
			
			\draw[black,<->] (-7,-1) -- (0,-1);
			\draw (5,-0.375) node[right=1pt] {\nc $ \eta_1$};
			\draw[black,<->] (5,-0.65) -- (5,-0.03);
			\draw (-3.2,0.25) node[right=1pt] {\nc $ \eta_2$};
			\draw[black,<->] (-3.2,0.47) -- (-3.2,0.03);
			\draw (-5,-1) node[below=1pt] {\nc $2R$};
				\draw[black,<->] (0,0.8) -- (3.5,0.8);
			\draw (2.5,0.8) node[above=1pt] {\nc $R$};
			\draw (-1.5,1.5) node[right=1pt] {\nc $E_0 + \ii \eta_0$};
					\node at (-1.5,1.5) {\textbullet};
						\draw (-1.5,-1.5) node[right=1pt] {\nc $E_0 - \ii \eta_0$};
					\node at (-1.5,-1.5) {\textbullet};
			
			\draw[black,thick,postaction={on each segment={mid arrow=black}}] (-3.5,0.5) arc[start angle=180, end angle=360, radius=3. 5];
			\draw[black,thick,postaction={on each segment={mid arrow=black}}] (3.5,0.5) -- (-3.5,0.5);
		\end{tikzpicture}
	\end{center}
	\caption{Sketch of the contours $\gamma_1$ (dashed) and $\gamma_2$ (full) from \eqref{eq:contdec1}--\eqref{eq:contdec2}. The union of the spectra of $H_1$ and $H_2$ is indicated in blue.}\label{fig:contours}
\end{figure}

The principal goal of this section is to prove \eqref{eq:echolong} in Theorem \ref{thm:echo1}~(ii), i.e.~study the behavior of $\Losa(t)$ defined in \eqref{eq:locechodef} for times $1\le t\lesssim \Delta^{-2}$.  In order to do so, we compute the random quantity $\left\langle \ee^{\ii tH_1}\Im G_1(E_0+\ii \eta_0)\ee^{- \ii tH_2}\right\rangle$ by residue calculus as
\begin{equation} \label{eq:rescalcstart}
	\begin{split}
		\left\langle \ee^{\ii tH_1}\Im G_1(E_0+\ii \eta_0)\ee^{- \ii tH_2}\right\rangle= & \left(\frac{1}{2\pi \ii}\right)^2 \oint_{\gamma_1}\oint_{\gamma_2}\ee^{\ii t(z_1-z_2)}\frac{\eta_0}{(z_1-E_0)^2 +\eta_0^2}\left\langle G_1(z_1)G_2(z_2)\right\rangle\dif z_1\dif z_2\\
		&\quad + \frac{1}{4\pi} \oint_{\gamma_2} \ee^{\ii t(E_0+i\eta_0-z_2)}\left\langle G_1(E_0+\ii \eta_0)G_2(z_2)\right\rangle\dif z_2.
	\end{split}
\end{equation}
Here, the contours $\gamma_1, \gamma_2$ are chosen to be two semicircles as indicated in Figure \ref{fig:contours}. More precisely, we take a (large) constant $R>0$ such that ${\rm supp}\rho_1$ and ${\rm supp}\rho_2$ are contained in $[-(R-1),R-1]$. The distance of the flat pieces from the real axis are denoted by $\eta_1:=\min\lbrace 1/t, \eta_0/2\rbrace$ and $0<\eta_2\lesssim 1/t$. The latter will explicitly be chosen later in Section \ref{subsec:contint}, where we conclude the proof of Theorem \ref{thm:echo1}~(ii). We decompose both contours into their flat in semicircular parts, $\gamma_j = \gamma_j^{(1)} \dot{+} \gamma_j^{(2)}$, $j \in [2]$, and parametrize them as follows: 
\begin{alignat}{4} \label{eq:contdec1}
\gamma_1^{(1)} &: \ z_1=E_1-\ii\eta_1 \quad &&\text{with} \quad E_1\in [-2R,2R]\,, \quad &&\gamma_1^{(2)} : \ z_1=2R\ee^{\ii \varphi}-\ii \eta_1 \quad &&\text{with} \quad \varphi\in [0,\pi] \\
\label{eq:contdec2}\gamma_2^{(1)} &: \ z_2=E_2+\ii\eta_2 \quad &&\text{with} \quad E_2\in [-R,R]\,, \quad &&\gamma_2^{(2)} : \ z_2=R\ee^{\ii \varphi}+\ii \eta_2 \quad &&\text{with} \quad \varphi\in [0,\pi]
\end{alignat}
Finally, we point out that, in order to \eqref{eq:rescalcstart} being valid, $\gamma_1$ is chosen in such a way that it encircles $E_0+\ii \eta_0$, but \emph{not} $E_0-\ii \eta_0$.

The following argument leading towards the proof of Theorem \ref{thm:echo1}~(ii) is split in three parts. First, in Section \ref{subsec:GL}, we approximate the random contour integrals \eqref{eq:rescalcstart} by their deterministic counterparts by using an appropriate \emph{two resolvent global law} for two different deformations (Proposition \ref{prop:GL}). Afterwards, in Section \ref{subsec:stabopshift}, we collect some preliminary stability bounds (Proposition \ref{prop:stab}) and information on the \emph{shift}, which is the key parameter in our analysis of the Loschmidt echo; see Lemmas \ref{lem:gen_shift}--\ref{lem:Im}. Finally, in Section~\ref{subsec:contint}, we summarize the evaluation of the deterministic contour integrals from Section \ref{subsec:GL} in five Lemmas \ref{lem:2nd_term}--\ref{lem:2nd_repl}. Combining these with estimates on the shift from Section \ref{subsec:stabopshift}, we conclude the proof of Theorem \ref{thm:echo1}~(ii) at the end of Section \ref{subsec:contint}.

\subsection{Step (i): Global law with two deformations} \label{subsec:GL}
The following two resolvent global law will be used to approximate \eqref{eq:rescalcstart} by its deterministic counterpart. 
\begin{proposition}[Average two resolvent global law] \label{prop:GL}
Let $D_1, D_2\in \C^{N \times N}$ be a bounded Hermitian matrices, i.e.~$\lVert D_j\rVert\le L$ for some $L > 0$, and $W$ a Wigner matrix satisfying Assumption \ref{ass:Wigner}. Moreover, let $z_1, z_2 \in \C$ be spectral parameters satisfying $\kappa := \min_{i \in [2]} \mathrm{dist}(z_i, [-(L+2), L+2]) \ge \delta > 0$ and denote $G_j(z_j) := (D_j + W - z_j)^{-1}$ for $j \in [2]$. Then it holds that 
\begin{equation} \label{eq:globallaw}
\left| \langle G_1(z_1) G_2(z_2) \rangle - \langle M(z_1, z_2)\rangle  \right| \prec \frac{C(\delta)}{N}\,, 
\end{equation}
where $C(\delta)> 0$ is a constant depending\footnote{By carefully tracking $\delta$ throughout the proof, one can see that the dependence is inverse polynomially, $C(\delta) \lesssim \delta^{-n}$ for some $n \in \N$. This will, however, be completely irrelevant for our purposes.} only on its argument (apart from $L$ and the constants from Assumption \ref{ass:Wigner}). In \eqref{eq:globallaw}, we abbreviated 
\begin{equation} \label{eq:2Gapprox}
M(z_1, z_2) = M_{12}(z_1, z_2):= \frac{M_1(z_1) M_2(z_2)}{1 - \langle M_1(z_1) M_2(z_2) \rangle}
\end{equation}
and $M_j = M_j(z_j)$, for $j \in [2]$, is the unique solution to the Matrix Dyson equation (MDE)
\begin{equation} \label{eq:MDE}
- \frac{1}{M_j} = z_j - D_j + \langle M_j \rangle \quad \text{with} \quad \Im M_j(z_j) \Im z_j > 0 \quad \text{for} \quad z_j \in \C\setminus \R \,. 
\end{equation}
\end{proposition}
\begin{proof}
	Using that $\Vert D_j + W \Vert \le L +2+ \epsilon$, $j \in [2]$ with very high probability and the stability bound $|1 - \langle M_1(z_1) M_2(z_2)\rangle|^{-1} \lesssim 1$ for $\kappa := \min_{i \in [2]} \mathrm{dist}(z_i, [-(L+2), L+2]) \gtrsim 1$ from Proposition~\ref{prop:stab} below,\footnote{In view of \eqref{eq:stabboundspecial}, note that the supports $\mathrm{supp}(\rho_1), \mathrm{supp}(\rho_2)$ of the scDos of $H_1$ and $H_2$ are contained in $[-(L+2), (L+2)]$.} the proof works in the same way as \cite[Proposition~3.1]{pretherm}, \cite[Appendix~B]{multiG}, \cite[Section~5.2]{iid}, or \cite[Section~6.2]{equipart}. We omit the details for brevity. 
\end{proof}

 Hence, by means of Proposition \ref{prop:GL}, we find that the random contour integral \eqref{eq:rescalcstart} can be approximated by the deterministic quantity
\begin{equation} \label{eq:detcontint}
	\begin{split}
		I_{E_0,\eta_0}(t):=&\left(\frac{1}{2\pi \ii}\right)^2 \oint_{\gamma_1}\oint_{\gamma_2}\ee^{\ii t(z_1-z_2)}\frac{\eta_0}{(z_1-E_0)^2 +\eta_0^2}\left\langle M(z_1,z_2)\right\rangle\dif z_1\dif z_2\\
		&\quad + \frac{1}{4\pi} \oint_{\gamma_2} \ee^{\ii t(E_0+i\eta_0-z_2)}\left\langle M(E_0+\ii \eta_0,z_2)\right\rangle\dif z_2.
	\end{split}
\end{equation}
up to an error of size $\mathcal{O}_\prec\big(C(\eta_1) C(\eta_2)/N\big)$, where we additionally used that the lengths of the contours are bounded, $\ell(\gamma_j) \lesssim 1$ for $j \in [2]$.

\subsection{Step (ii): Preliminary bounds on the stability operator and the shift} \label{subsec:stabopshift}
As usual in random matrix theory, local/global laws are governed by a \emph{stability operator}, which, in our case is given by 
\begin{equation} \label{eq:stabopdef}
\mathcal{B}_{12}(z_1, z_2)[\cdot] := \mathbf{1} - M_1 \langle \cdot \rangle M_2 \quad \text{with}\quad M_j \equiv M_j(z_j)\,. 
\end{equation}
One can easily see that $\mathcal{B}_{12}(z_1, z_2)$ has a highly degenerate eigenvalue one, and its only non-trivial eigenvalue is given by $1 - \langle M_1 M_2\rangle$ with corresponding eigen``vector" $M_1 M_2$.  

The following proposition, whose proof is given in Section \ref{sec:stabopshiftpf}, states an upper bound on the inverse of this non-trivial eigenvalue. A simplified form this stability bound already appeared in \cite[Lemma~5.2]{quenchuniv} for the very special case that $D_1 = \alpha D_2$ for some $\alpha \in \R$.

\begin{proposition}[Stability bound]\label{prop:stab} Fix a (large) $L>0$. Uniformly in $z_1,z_2\in\C\setminus\R$ and traceless Hermitian $D_1,D_2$ with $\vert z_j\vert\le L$, $\lVert D_j\rVert\le L$, $j=1,2$, it holds that
	\begin{equation}
		\left\vert \frac{1}{1-\langle M_1M_2\rangle}\right\vert\lesssim \frac{1}{\Delta^2 + \left( \Re z_1-\Re z_2\right)^2 + \left(\Im \langle M_1\rangle + \Im \langle M_2\rangle\right)^2+\left\vert\frac{ \Im z_1}{\langle\Im M_1\rangle}\right\vert+ \left\vert\frac{ \Im z_2}{\langle\Im M_2\rangle}\right\vert}\vee 1\,, 
		\label{eq:stab_bound}
	\end{equation}
	where we denoted $\Delta^2:=\langle (D_1-D_2)^2\rangle$. 
\end{proposition}

In the current Section \ref{sec:pfecho1long}, more precisely, the proof of Proposition \ref{prop:GL} above, only the special case 
\begin{equation} \label{eq:stabboundspecial}
	\left\vert 1-\langle M_1M_2\rangle\right\vert^{-1} \lesssim 1 \quad \text{for} \quad \max_{j \in [2]} \mathrm{dist}(z_j, \mathrm{supp}(\rho_j)) \gtrsim 1
\end{equation}
of Proposition \ref{prop:stab} is relevant. However, for later reference, we also point out that, in particular, $\vert 1-\langle M_1M_2\rangle\vert^{-1}\lesssim \vert z_1-z_2\vert^{-2}$ and that the lhs.~of \eqref{eq:stab_bound} is bounded by one, whenever $z_1,z_2$ are in the same half-plane and $\rho_1(z_1) + \rho_2(z_2) \gtrsim 1$ (e.g.~if one of them is in the bulk, $\Re z_j \in \mathbf{B}_\kappa(\rho_j)$).

In addition to these bounds, Proposition \ref{prop:stab} also plays an important role in the analysis of the \emph{shift} $\mathfrak{s}(z_1,z_2)$ of the spectral parameters $z_1, z_2$ in the \emph{(generalized) $M$-resolvent identity} 
\begin{equation}
\langle M_{12}\rangle = 	\frac{\langle M_1M_2\rangle}{1-\langle M_1M_2\rangle} = \frac{\langle M_1\rangle - \langle M_2\rangle}{z_1-z_2-\mathfrak{s}(z_1,z_2)}, 
	\label{eq:identity}
\end{equation} 
which can easily be obtained by subtracting MDEs \eqref{eq:MDE} for $M_2$ and $M_1$ from each other. In \eqref{eq:identity}, the shift is defined as follows. 

\begin{definition}[The \emph{shift}] Let $D_1, D_2$ be Hermitian traceless matrices and let $M_j(z_j)$ for $j \in [2]$ be the solution of the MDE \eqref{eq:MDE}. Then, we define the \emph{shift} (depending on $D_1, D_2$ and $z_1, z_2 \in \C\setminus \R$) as
	\begin{equation}
		\mathfrak{s}(z_1,z_2):=\frac{\langle M_1(z_1)(D_1-D_2)M_2(z_2)\rangle}{\langle M_1(z_1)M_2(z_2)\rangle}\,, 
		\label{eq:shift}
	\end{equation}
	whenever the denominator does not vanish. 
\end{definition}

As already mentioned above, the shift $\mathfrak{s}$ is the key parameter in our analysis of the Loschmidt echo. We now collect several estimates on $\mathfrak{s}$ in the following Lemmas \ref{lem:gen_shift}--\ref{lem:Im}. The proofs, which are based on the stability bound in Proposition \ref{prop:stab}, are given in Section \ref{sec:stabopshiftpf}. 

\begin{lemma}[Properties of $\ms(z_1,z_2)$]\label{lem:gen_shift}
	Fix a (small) $\kappa>0$ and a (large) $L>0$. Consider spectral parameters $z_1,z_2 \in \C\setminus \R$ such that $\Im z_1 \Im z_2<0$ and $\vert z_j\vert\le L$, $\lVert D_j\rVert\le L$, for $j \in [2]$. Assume that at least one of these parameters is such that the (positive) harmonic extension of the scDos is positive, i.e.~$\rho_1(z_1)+\rho_2(z_2)\ge \kappa$. Then there exists a positive constant $\mathfrak{c}$ which depends only on $\kappa, L$ such that for any Hermitian traceless $D_1,D_2$ with $\Delta := \langle (D_1 -D_2)^2 \rangle^{1/2}\le \mc$ we have the following: 
	\begin{enumerate}
		\item The denominator of the shift \eqref{eq:shift} is of order one, $\vert\langle M_1(z_1)M_2(z_2)\rangle\vert\sim 1$. In particular,
		\begin{equation}
			\vert\mathfrak{s}(z_1,z_2)\vert\lesssim \Delta.
			\label{eq:s_bound}
		\end{equation}
		\item If $\rho_j(z_j) \ge \kappa/2$, then
		\begin{equation}
			\vert\partial_{z_j} \mathfrak{s}(z_1,z_2)\vert\lesssim \Delta.
			\label{eq:s_deriv_bound}
		\end{equation}
	\end{enumerate}
	Here all implicit constants depend only on $\kappa$ and $L$.
\end{lemma}

We now introduce an auxiliary function $f$, which exactly detects the influence of the shift on the real part of a spectral parameter. 

\begin{lemma}[Definition of $f$ and $\ms_0$] \label{lem:f}
	Fix a (small) $\kappa>0$ and a (large) $L>0$. Consider $0 < \eta_1, \eta_2<L$ and a spectral parameter $z_2=E_2+\ii\eta_2$ such that $\rho_2(z_2)\ge \kappa$, and satisfying $\vert z_2\vert\le L$. Let $D_1, D_2$ be Hermitian traceless matrices with $\lVert D_j\rVert\le L$, $j\in [2]$. Assume that $\Delta := \langle (D_1 -D_2)^2 \rangle^{1/2} \le\mc$, where $\mc$ is the constant from Lemma \ref{lem:gen_shift}. 
	
	Then there exists a unique \emph{energy renormalization} $f^{\eta_1,\eta_2}(E_2)= f(E_2)\in\mathbb{R}$ with $\vert f(E_2)\vert\le L$ such that
	\begin{equation*}
		\Re\left(f(E_2)-E_2-\ms(f(E_2)-\ii\eta_1,E_2+\ii\eta_2))\right)=0.
	\end{equation*}
	
	Moreover, denoting the \emph{renormalized (one point) shift} by 
\begin{equation}\label{s0def}
	\ms_0^{\eta_1,\eta_2}(E_2):=\ms (f(E_2)-\ii\eta_1,E_2+\ii\eta_2),
\end{equation}
	 the functions $f^{\eta_1,\eta_2}(E_2)$ and $\ms_0^{\eta_1,\eta_2}(E_2)$ are differentiable in $\eta_1,\eta_2$ and for $E_2 \in \mathbf{B}_\kappa(\rho_2)$ in the bulk, and the derivatives satisfy 
	\begin{equation}
		\vert\partial_{E_2}f^{\eta_1,\eta_2}(E_2) - 1\vert\lesssim\Delta,\quad \left\vert\partial_{\eta_j} f^{\eta_1,\eta_2}(E_2)\right\vert \lesssim \Delta,\,j\in [2],\quad  \text{and} \quad  \vert\partial_{E_2}\ms_0^{\eta_1,\eta_2}(E_2)\vert\lesssim\Delta.
		\label{eq:deriv_bounds}
	\end{equation}
\end{lemma}
Whenever it does not lead to confusion our ambiguities, we will omit the superscripts $\eta_1,\eta_2$ of $f^{\eta_1,\eta_2}$ and $\ms_0^{\eta_1,\eta_2}$. Next, we show that the \emph{imaginary part} of the renormalized shift is in fact much smaller than indicated by the upper bounds of order $\Delta$ in \eqref{eq:s_bound}--\eqref{eq:s_deriv_bound} and \eqref{eq:deriv_bounds}.

\begin{lemma}[Behavior of $\Im\ms_0$] \label{lem:Im} Fix a (small) $\kappa>0$ and a (large) $L>0$. Let $E \in \mathbf{B}_\kappa(\rho_2) $ be in the bulk of $\rho_2$. Then there exist positive constants $c_1, c_2 > 0$ such that for any Hermitian traceless $D_1, D_2$ with $\lVert D_j\rVert\le L$, $j=1,2$, $\Delta<c_1$ and for any $0<\eta_j\le c_2\Delta$, for $j \in [2]$, it holds that
	\begin{equation}
		\Im\ms_0^{\eta_1,\eta_2}(E)\sim \Delta^2.
		\label{eq:Ims}
	\end{equation}
	Here, $c_1,c_2$ and the implicit constants in \eqref{eq:Ims} depend only on $\kappa$ and $L$.
\end{lemma}

In the following section, armed with the preliminary bounds from Proposition \ref{prop:stab} and Lemmas \ref{lem:gen_shift}--\ref{lem:Im}, we carry out the evaluation of the contour integrals in \eqref{eq:detcontint}.

\subsection{Step (iii): Contour integration of the deterministic approximation} \label{subsec:contint}
Throughout this section, let $[a,b]$ be an interval with length of order one satisfying $\mathrm{dist}(E_0, [a,b]^c) \gtrsim 1$ and $ \mathrm{dist}([a,b], (\mathrm{supp}(\rho_1) \cap \mathrm{supp}(\rho_2))^c) \gtrsim 1$. That is,  the energy $E_0$ from Theorem \ref{thm:echo1} is order one away from the boundary of $[a,b]$ and $[a,b]$ is simultaneously in the bulk of $\rho_1$ and $\rho_2$. The existence of such an interval is always guaranteed.

As already mentioned above, we now dissect the evaluation of \eqref{eq:detcontint} in several parts. As the first step, we show that the second line of \eqref{eq:detcontint} is in fact negligible. The proofs of Lemma \ref{lem:2nd_term} and all the other Lemmas~\ref{lem:tails}--\ref{lem:2nd_repl} is given in Section \ref{sec:conintproof}.
\begin{lemma}[The second line is negligible]\label{lem:2nd_term}
	Under the assumptions of Theorem \ref{thm:echo1}~(ii) it holds that
	\begin{equation*}
		I_{E_0}^{(2)}:= \frac{1}{4\pi} \oint_{\gamma_2} \ee^{\ii t(E_0+\ii \eta_0-z_2)}\left\langle M(E_0+i\eta_0,z_2)\right\rangle\dif z_2 = \mathcal{O}\left(\frac{1}{t}\right).
	\end{equation*} 
\end{lemma}

For the remaining first line of \eqref{eq:detcontint}, we then find that the main contribution of the $\gamma_2$ integral comes from the interval $[a,b] + \ii \eta_2$, i.e.~we can cut away the tails. 

\begin{lemma}[Cutting tails]\label{lem:tails} 	Under the assumptions of Theorem \ref{thm:echo1}~(ii) it holds that
	\begin{equation*}
		\begin{split}
			I_{E_0}^{(1)}:=&\left(\frac{1}{2\pi \ii}\right)^2 \oint_{\gamma_1}\oint_{\gamma_2}\ee^{\ii t(z_1-z_2)}\frac{\eta_0}{(z_1-E_0)^2 +\eta_0^2}\langle M(z_1,z_2)\rangle\dif z_1\dif z_2\\
			= &\left(\frac{1}{2\pi \ii}\right)^2 \oint_{\gamma_1}\int_a^b \ee^{\ii t(z_1-E_2-i\eta_2)}\frac{\eta_0}{(z_1-E_0)^2 +\eta_0^2}\langle M(z_1,E_2+\ii\eta_2)\rangle\dif z_1\dif E_2 + \mathcal{O}\left(\frac{1}{t}+\frac{\eta_0}{\Delta}\right).
		\end{split}
	\end{equation*}
\end{lemma}

The following lemma formally implements inside the integral from Lemma \ref{lem:tails} the approximation 
\begin{equation*}
\langle M(z_1, E_2 + \ii \eta_2)\rangle = 	 \frac{\langle M_1(z_1)\rangle - \langle M_2(E_2 + \ii \eta_2)\rangle}{z_1-(E_2 + \ii \eta_2)-\mathfrak{s}(z_1,E_2 + \ii \eta_2)} \approx \frac{\langle M_1(z_1)\rangle - \langle M_2(E_2+\ii\eta_2)\rangle}{z_1-(E_2+\ii\eta_2)-\ms_0^{\eta_1,\eta_2}(E_2)}\,, 
\end{equation*}
which is valid in the main contributing regime $E_1\approx E_2$. This is our \emph{first replacement} $\mathfrak{s}(z_1,E_2 + \ii \eta_2) \to \ms_0^{\eta_1,\eta_2}(E_2)$.

\begin{lemma}[First replacement]\label{lem:1st_repl} Denote $\mathfrak{d}:=\min_{E_2\in [a,b]}\vert \eta_1+\eta_2+\Im \ms_0^{\eta_1, \eta_2}(E_2)\vert$. 
Then, under the assumptions of Theorem \ref{thm:echo1}~(ii), it holds that
	\begin{equation} \label{eq:firstrepl}
		\begin{split}
			&\left(\frac{1}{2\pi \ii}\right)^2 \oint_{\gamma_1}\dif z_1\int_a^b \dif E_2 \, \ee^{\ii t(z_1-E_2-\ii \eta_2)}\frac{\eta_0}{(z_1-E_0)^2 +\eta_0^2}\langle M(z_1,E_2+\ii\eta_2)\rangle\\
			= &\left(\frac{1}{2\pi \ii}\right)^2 \oint_{\gamma_1} \dif z_1\int_a^b \dif E_2 \, \ee^{\ii t(z_1-E_2-\ii\eta_2)}\frac{\eta_0}{(z_1-E_0)^2 +\eta_0^2}\cdot \frac{\langle M_1(z_1)\rangle - \langle M_2(E_2+\ii\eta_2)\rangle}{z_1-(E_2+\ii \eta_2)-\ms_0^{\eta_1,\eta_2}(E_2)} \\
			&\qquad + \mathcal{O}\left(\eta_0+\Delta\vert\log\Delta\vert+\Delta\vert\log\mathfrak{d}\vert\right).
		\end{split}
	\end{equation}
\end{lemma}
Next, plugging in the Stieltjes representation $\langle M_1(z_1)\rangle = \int_{\R}\rho_1(x)(x-z_1)^{-1}\dif x$, the $\gamma_1$ integral in Lemma \ref{lem:1st_repl} can be explicitly computed using residue calculus. The ``unwanted" residue contributions arising in this way can be estimated using the oscillatory factor and integration by parts (see the proof of Lemma~\ref{lem:1st_comp} in Section \ref{sec:conintproof}). 
\begin{lemma}[Residue computation after the first replacement]\label{lem:1st_comp} Denote $\mathfrak{a}:=\min_{E_2\in [a,b]}\vert\eta_0-\eta_2- \Im\ms_0^{\eta_1,\eta_2}(E_2)\vert$ and suppose that
	\begin{equation}
		\eta_1+\eta_2+\Im \ms_0(E_2)>0,\quad \forall E_2\in [a,b]\,. 
		\label{eq:1_condition}
	\end{equation}
Then, again under the assumptions of Theorem \ref{thm:echo1}~(ii), it holds that
	\begin{equation} \label{eq:rescompfirst}
		\begin{split}
			&\left(\frac{1}{2\pi \ii}\right)^2 \oint_{\gamma_1}\dif z_1\int_a^b \dif E_2\ee^{\ii t(z_1-E_2-i\eta_2)}\frac{\eta_0}{(z_1-E_0)^2 +\eta_0^2}\cdot \frac{\langle M_1(z_1)\rangle - \langle M_2(E_2+\ii\eta_2)\rangle}{z_1-(E_2+\ii \eta_2)-\ms_0^{\eta_1,\eta_2}(E_2)} \\
			 = &-\frac{1}{2\pi \ii}\int_{\R}\dif x\int_{a}^b\dif E_2 \ee^{\ii t(x-E_2-\ii \eta_2)}\frac{\eta_0}{(x-E_0)^2+\eta_0^2}\cdot\frac{\rho_1(x)}{x-(E_2+\ii \eta_2)-\ms_0^{\eta_1,\eta_2}(E_2)}\\
			&\qquad + \mathcal{O}\left(\frac{\vert\log\mathfrak{a}\vert}{t} +\frac{\Delta + t^{-1}}{t\mathfrak{a}}+\eta_0\vert\log\mathfrak{a}\vert+\frac{\eta_0(\Delta+ t^{-1})}{\mathfrak{a}}\right).
		\end{split}
	\end{equation}
\end{lemma}

In the following lemma, we (i) complete the integral $\int_a^b$ to a full contour integral $\oint_{\gamma_2}$, i.e.~put back the tails that were cut away in Lemma \ref{lem:tails}, and (ii) implement the \emph{second replacement}
\begin{equation}
\label{eq:2ndrepl} \mathfrak{s}_0^{\eta_1,  \eta_2}(E_2) \to \mathfrak{s}_0 := \mathfrak{s}_0^{\eta_1,  \eta_2}\left(\left(f^{\eta_1,\eta_2}\right)^{-1}(E_0)\right)
\end{equation}
 inside the integral from Lemma \ref{lem:1st_comp}. This replacement leads to a small error comparing to the leading term since $\mathfrak{s}_0^{\eta_1,  \eta_2}(E_2)\approx \ms_0$ in the relevant regime $E_2\approx E_0$.

\begin{lemma}[Second replacement]\label{lem:2nd_repl} Let $\mathfrak{b}:=\min_{E_2\in [a,b]}\vert \eta_2+\Im \ms_0^{\eta_1,\eta_2}(E_2)\vert$ and $\ms_0$ as in \eqref{eq:2ndrepl}. 
Then, again under the assumptions of Theorem \ref{thm:echo1}~(ii), it holds that
	\begin{equation}
		\begin{split}
			& -\frac{1}{2\pi \ii }\int_{\R} \dif x\int_{a}^b \dif E_2 \, \ee^{\ii t(x-E_2-\ii\eta_2)}\frac{\eta_0}{(x-E_0)^2+\eta_0^2}\cdot\frac{\rho_1(x)}{x-(E_2+i\eta_2+\ms_0^{\eta_1, \eta_2}(E_2))}\\
			= &-\frac{1}{2\pi \ii}\int_{\R}\dif x\oint_{\gamma_2} \dif z_2 \ee^{\ii t(x-z_2)}\frac{\eta_0}{(x-E_0)^2+\eta_0^2}\cdot\frac{\rho_1(x)}{x-(z_2+\ms_0)}\\
			&\qquad + \mathcal{O}\left(\frac{\eta_0+\mb}{\mb}\Delta\vert\log(\eta_0+\mb)\vert +\eta_0 \vert\log\mb\vert+\frac{1}{t}\right).
		\end{split}
		\label{eq:2nd_repl}
	\end{equation}
\end{lemma}

Armed with Lemmas \ref{lem:2nd_term}--\ref{lem:2nd_repl}, we can finally give the proof of Theorem \ref{thm:echo1}~(ii). 

\begin{proof}[Proof of Theorem \ref{thm:echo1}~(ii)]
Combining Lemmas \ref{lem:2nd_term} - \ref{lem:2nd_repl} we find that
\begin{equation}
	I_{E_0,\eta_0}(t) = -\frac{1}{2\pi \ii}\int_{\R}\oint_{\gamma_2} \ee^{\ii t(x-z_2)}\frac{\eta_0}{(x-E_0)^2+\eta_0^2}\cdot\frac{\rho_1(x)\dif x}{x-(z_2+\ms_0)}\dif z_2 + \mathcal{O}\big(\widehat{\mathcal{E}}(t)\big),
	\label{eq:I_final}
\end{equation}
where we collected all the error terms in
\begin{equation*}
	\widehat{\mathcal{E}}(t):=\frac{\eta_0}{\Delta}+\Delta\vert\log\Delta\vert + \Delta\vert \log \mathfrak{d}\vert + \frac{\vert\log\mathfrak{a}\vert}{t} +\frac{\Delta + t^{-1}}{t\mathfrak{a}}+\eta_0\vert\log\mathfrak{a}\vert+\frac{\eta_0(\Delta+ t^{-1})}{\mathfrak{a}} + \frac{\eta_0+\mb}{\mb}\Delta\vert\log(\eta_0+\mb)\vert +\eta_0 \vert\log\mb\vert\,.
\end{equation*}

We shall now estimate $\widehat{\mathcal{E}}(t)$ in different time regimes. First note that Lemmas \ref{lem:gen_shift} and \ref{lem:Im} imply the existence of positive constants $\lbrace c_j\rbrace_{j=1}^4$ such that
\begin{equation}
	\begin{split}
		\vert \ms(z_1,z_2)\vert \le c_1\Delta,\quad &\text{for all} \quad  \vert z_1\vert\le 2R,\, E_2\in [a,b],\,\eta_2\in[0,1], \quad \text{and}\\
		c_2\Delta^2\le \Im\ms_0^{\eta_1,\eta_2}(E_2)\le c_3\Delta^2,\quad &\text{for all} \quad  E_2\in[a,b],\,\eta_j\in [0,c_4\Delta], j=1,2.
	\end{split}
	\label{eq:constants}
\end{equation}
\underline{First regime:} For $1\le t\le 4Kc_3/(c_4 \Delta)$ we take $\eta_2:=8Kc_1c_3/(c_4t)$. Then, for any $E_2\in [a,b]$ it holds that
\begin{equation*}
	\eta_2+\Im\ms_0(E_2)\ge 8Kc_1c_3/(c_1t) - c_1\Delta\ge 4Kc_1c_3/(c_1t)>0.
\end{equation*}
In particular, the parameters $\mathfrak{a}$,  $\mathfrak{b}$, and $\mathfrak{d}$ from Lemmas \ref{lem:1st_comp}, \ref{lem:2nd_repl}, and \ref{lem:1st_repl}, respectively, are all of order $1/t$ and $\widehat{\mathcal{E}}(t)$ is bounded as
\begin{equation}
	\widehat{\mathcal{E}}(t)\lesssim \frac{1+\log t}{t}+\frac{\eta_0}{\Delta}+\Delta\vert\log \Delta\vert+\Delta\log t,\quad \text{for} \quad  1\le t\le \frac{4Kc_3}{c_4}\cdot\frac{1}{\Delta}.
	\label{eq:t<1/Delta}
\end{equation}
\underline{Second regime:} For $4Kc_3/(c_4 \Delta)\le t\le 2Kc_3/\eta_0$, we take $\eta_2:=\frac{4Kc_3}{t}$. In this regime, $\eta_2\le c_4\Delta$, so the positivity of $\eta_2+\Im\ms_0(E_2)$ follows from \eqref{eq:constants}. We also have that $\eta_2\ge 2\eta_0$ and again $\mathfrak{a}\sim \mathfrak{b}\sim \mathfrak{d}\sim 1/t$. Therefore, \eqref{eq:t<1/Delta} holds in the whole regime $1\le t\le 2Kc_3/\eta_0$. \\[2mm]
\noindent\underline{Third regime:} It remains to study the regime $2Kc_3/\eta_0 \le t \le K/\Delta^2$.  If $\eta_0\le 2c_3\Delta^2$ it is in fact empty, hence we may assume $\eta_0\ge 2c_3\Delta^2$. In this case, we take $\eta_2:=\min\lbrace \eta_0/4,c_4\Delta,1/t\rbrace$ and find that $\mathfrak{a}\sim \eta_0$, $\mathfrak{b}\gtrsim\Delta^2$, $\mathfrak{d}\gtrsim\Delta^2$. Moreover, the error term $\widehat{\mathcal{E}}(t)$ is bounded as
\begin{equation*}
	\widehat{\mathcal{E}}(t)\lesssim \frac{1+\log t}{t}+\Delta\vert\log \Delta\vert+\frac{\eta_0\vert\log\Delta\vert}{\Delta},\quad \text{for} \quad  2Kc_3/\eta_0\le t\le K/\Delta^2.
\end{equation*}

After having chosen $\eta_2$ in all time regimes explicitly, we can perform $z_2$-integration in \eqref{eq:I_final}. Note that in all time regimes $\eta_2$ was chosen in such a way that $\eta_2+\Im \ms_0>0$, which guarantees that $\gamma_2$ encircles the point $x-\ms_0$ for $x\in{\rm supp}(\rho_1)$. So, \eqref{eq:I_final} evaluates to
\begin{equation}\label{Iphase}
	\begin{split}
		&I_{E_0,\eta_0}(t)=\ee^{\ii t\ms_0}\int_{\R}\frac{\eta_0}{(x-E_0)^2+\eta_0^2}\rho_1(x)\dif x +\mathcal{O}\big(\widehat{\mathcal{E}}(t)\big) = \ee^{\ii t\ms_0}\Im \langle M_1(E_0+\ii\eta_0)\rangle + \mathcal{O}\big(\widehat{\mathcal{E}}(t)\big).
	\end{split}
\end{equation}
After dividing by $\Im \langle M_1(E_0+\ii \eta_0)\rangle $ and taking the absolute value square, it is left to notice that, setting 
\begin{equation} \label{eq:Gamma}
\Gamma := 2 \, \Im\ms_0^{0,0}\left(\left(f^{0,0}\right)^{-1}(E_0)\right)\,, 
\end{equation}
it holds that
\begin{equation*}
	\Im\ms_0=\Im\ms_0^{\eta_1,\eta_2}\left(\left(f^{\eta_1,\eta_2}\right)^{-1}(E_0)\right) = \Gamma/2 +\mathcal{O}(\Delta(\eta_1+\eta_2))= \Gamma/2+\mathcal{O}(\Delta/t).
\end{equation*}
Here we used \eqref{eq:deriv_bounds} from Lemma \ref{lem:f} and \eqref{eq:s_deriv_bound} from Lemma \ref{lem:gen_shift} together with the bound $\eta_j\lesssim 1/t$, $j=1,2$. By Lemma \ref{lem:Im}, we finally see that the implicit constants in $\Gamma \sim \Delta^2$ only depend on $\kappa$ and $L$. This finishes the proof of Theorem \ref{thm:echo1}~(ii).
\end{proof}

\section{Stability operator and shift: Proofs for Section \ref{subsec:stabopshift}} \label{sec:stabopshiftpf}

\subsection{Bound on the stability operator: Proof of Proposition \ref{prop:stab}}
Throughout the proof, we will use the shorthand notations $E_j:=\Re z_j$, $\eta_j:=\vert\Im z_j\vert$, $\rho_j:=\frac{1}{\pi}\left\vert\langle\Im M_j(z_j)\rangle\right\vert$ and $\omega_j:=z_j+\langle M_j(z_j)\rangle$, for $j\in [2]$. 

We will conclude Proposition \ref{prop:stab} from the following lemma. 

\begin{lemma} \label{lem:proofstab}
Under the assumptions of Proposition \ref{prop:stab} and using the notations from above, we have that: 
\begin{align}
	\vert 1-\langle M_1M_2\rangle\vert^{-1}&\lesssim (\eta_1/\rho_1+\eta_1/\rho_2)^{-1}\vee 1.
\label{eq:stab_Im} \\
	\vert 1-\langle M_1M_2\rangle\vert^{-1}&\lesssim (\Delta^2 + \vert\omega_1-\overline{\omega}_2\vert^2)^{-1}.
\label{eq:stab_D} \\
|1 - \langle M_1 M_2 \rangle|^{-1} &\lesssim |z_1 - z_2|^{-2} 
	\label{eq:stab_E}
\end{align}
\end{lemma}
Combining \eqref{eq:stab_Im}--\eqref{eq:stab_E} with the simple observation $\vert \omega_1-\overline{\omega}_2\vert\ge \vert\langle\Im M_1\rangle+\langle\Im M_2\rangle\vert$, we conclude \eqref{eq:stab_bound}, i.e.~the proof of Proposition \ref{prop:stab}. \qed

\begin{proof}[Proof of Lemma \ref{lem:proofstab}]
For \eqref{eq:stab_Im}, it is sufficient to check that for some $c\in (0,1)$ we have $\vert\langle M_1M_2\rangle\vert \le \left(1-c(\eta_1/\rho_1+\eta_2/\rho_2)\right)\vee (1-c)$. This follows from a simple Cauchy-Schwarz inequality $\vert \langle M_1M_2\rangle\vert\le \langle |M_1|^2\rangle^{1/2} \langle |M_2|^2\rangle^{1/2}$ together with the estimate 
\begin{equation*}
\langle |M_j|^2\rangle^{1/2}=\left(\frac{\langle \Im M_j\rangle}{\Im z_j+\langle\Im M_j\rangle}\right)^{1/2}\lesssim \left(\frac{\rho_j}{\eta_j+\rho_j}\right)^{1/2}\le \left(1-\frac{1}{2}\cdot \frac{\eta_j}{\rho_j}\right)\vee (1-c)\,, \quad j \in [2]
\end{equation*}
where the first step follows by taking the imaginary part of the MDE \eqref{eq:MDE}. 

For \eqref{eq:stab_D}, we note that it is sufficient to show
\begin{equation}
	\Re \langle M_1M_2\rangle \le 1-c\left(\left\langle (D_1-D_2)^2\right\rangle+\vert \omega_1-\bar{\omega}_2\vert^2\right) \quad \text{for some} \quad c > 0\,. 
	\label{eq:stab_D_aux}
\end{equation}
The idea for proving \eqref{eq:stab_D_aux} is to translate it to a question for the spectral measures of $D_1$ and $D_2$.

In order to do so, for $j \in [2]$, denote the eigenvalues and eigenvectors of $D_j$ by $\{\lambda_k^{(j)}\}_{k=1}^N$ and $\lbrace\bm{u}_k^{(j)}\rbrace_{k=1}^N$, respectively, and the normalized spectral measure by $\mu_j:=N^{-1}\sum_{k=1}^N \delta_{\lambda_k^{(j)}}$. By the MDE \eqref{eq:MDE}, we immediately see that $\omega_j$ solves the equation $\omega_j-z_j=m_{\mu_j}(\omega_j)$, where $m_\mu(z) := \int_\R \dif \mu(x) (x-z)^{-1}$ is the Stieltjes transform of the probability measure $\mu$. By taking the imaginary part and estimating $\vert \Im\omega_j\vert>\vert \Im \omega_j-\Im z_j\vert$ we hence find
\begin{equation}
	\int\frac{\dif \mu_j(x)}{\vert x-\omega_j\vert^2}<1\,.
	\label{eq:omega_int_bound}
\end{equation}

Using the above notations, we further see that $M_j$ can be written as 	$M_j=\sum_{k=1}^N (\lambda_k^{(j)}-\omega_j)^{-1}\vert \bm{u}_k^{(j)}\rangle \langle \bm{u}_k^{(j)}\vert$ and thus
\begin{equation*}
	\langle M_1M_2\rangle =\frac{1}{N^2}\sum_{a,b=1}^{N}\frac{1}{\lambda_a^{(1)}-\omega_1}\cdot \frac{1}{\lambda_b^{(2)}-\omega_2}	f\big(\lambda_a^{(1)},\lambda_b^{(2)}\big)\,, \quad \text{with} \quad 	f\big(\lambda_a^{(1)},\lambda_b^{(2)}\big):=N\big\vert\big\langle \bm{u}_a^{(1)},  \bm{u}_b^{(2)}\big\rangle\big\vert^2\,. 
\end{equation*}
Extending $f(x,y)$ to $\R^2$ by zero, we immediately see the following properties of $f$:
\begin{enumerate}
	\item $f(x,y)\ge 0$ for all $x,y\in\mathbb{R}$.
	\item $\int f(x,y)\dif \mu_2(y)=\mathbf{1}_{{\rm supp}\, \mu_1}(x)$ and $\int f(x,y)\dif \mu_1(x) = \mathbf{1}_{{\rm supp}\, \mu_2}(y)$.
	\item On $\R^2$, $\dif \nu(x,y):=f(x,y)\dif \mu_1(x)\dif \mu_2(y)$ is a probability measure with marginals $\mu_1$ and $\mu_2$.
\end{enumerate}

In this way, the desired inequality \eqref{eq:stab_D_aux} can equivalently be rewritten as
\begin{equation}
	\Re\iint \frac{1}{x-\omega_1}\cdot\frac{1}{y-\omega_2}\dif \nu(x,y) \le 1- c\left(\iint (x-y)^2\dif\nu(x,y)+\vert\omega_1-\bar{\omega}_2\vert^2\right)\,. 
	\label{eq:stab_D_nu}
\end{equation}
In this form, using \eqref{eq:omega_int_bound}, we begin by estimating the lhs.~of \eqref{eq:stab_D_nu} as
\begin{equation*}
	\Re\iint \frac{1}{x-\omega_1}\cdot\frac{1}{y-\omega_2}\dif \nu(x,y) 
		<1-\frac{1}{2}\iint\left\vert \frac{1}{x-\omega_1}-\frac{1}{y-\bar{\omega}_2}\right\vert^2\dif \nu(x,y),
\end{equation*}
Thus, in order to arrive at \eqref{eq:stab_D_nu}, it suffices to bound 
\begin{equation*}
	\begin{split}
		&\iint\left\vert \frac{1}{x-\omega_1}-\frac{1}{y-\bar{\omega}_2}\right\vert^2\dif \nu(x,y)\gtrsim \iint \left\vert (x- y)-(\omega_1-\bar{\omega}_2)\right\vert^2\dif \nu(x,y)\\
		&\quad = \iint (x-y)^2\dif\nu(x,y)-2\Re (\omega_1-\bar{\omega}_2)\iint (x-y)\dif \nu(x,y)+\vert \omega_1-\bar{\omega}_2\vert^2\\
		&\quad=\iint (x-y)^2\dif\nu(x,y)+\vert \omega_1-\bar{\omega}_2\vert^2.
	\end{split}
\end{equation*}
where in the first step we employed $\vert x-\omega_1\vert \lesssim \lVert D_1\rVert +\vert \omega_1\vert\lesssim 1$ (and analogously for $\vert y-\bar{\omega}_2\vert$), while in the last step we used that fact that $D_1$ and $D_2$ are traceless. This finishes the proof of \eqref{eq:stab_D}. 

Finally, for \eqref{eq:stab_E}, we use  \eqref{eq:stab_D} and \eqref{eq:identity} to get that
\begin{equation}
	\begin{split}
	\vert z_1-z_2\vert^2 &=\left\vert \langle M_2(D_1-D_2)M_1\rangle + (1-\langle M_1M_2\rangle)(z_1-z_2+\langle M_1\rangle - \langle M_2\rangle)\right\vert^2\\
		& \lesssim  \vert\langle M_1(D_1-D_2)M_2\rangle\vert^2 + \vert 1-\langle M_1M_2\rangle\vert \\
		&\lesssim \left\langle (D_1-D_2)^2\right\rangle + \vert 1-\langle M_1M_2\rangle\vert \lesssim \vert 1-\langle M_1M_2\rangle\vert\,. \qedhere
	\end{split}
	\label{eq:stab_E_aux}
\end{equation}
\end{proof}

\subsection{Properties of the shift: Proof of Lemmas \ref{lem:gen_shift}--\ref{lem:Im}}
We finally prove the properties of the shift from Lemmas \ref{lem:gen_shift}--\ref{lem:Im}.

\begin{proof}[Proof of Lemma \ref{lem:gen_shift}] The proof is split in two parts in the statement of the lemma. 
	\\[2mm]
	\underline{Part (1):}
	Given $\vert\langle M_1M_2\rangle\vert\sim 1$, note that the bound \eqref{eq:s_bound} immediately follows since, if, say, $z_1$ is such that $\rho_1(z_1) \ge \kappa/2$, then $\lVert M_1\rVert\lesssim 1$ and $\langle |M_2|^2\rangle^{1/2}\le 1$. Both of these estimates easily follow by taking the imaginary part of the respective MDEs \eqref{eq:MDE}. 
	
	It is hence left to prove $\vert\langle M_1M_2\rangle\vert\sim 1$. The upper bound $\vert\langle M_1M_2\rangle\vert\le 1$ is a consequence of the Cauchy-Schwarz inequality and $\langle |M_j|^2\rangle^{1/2}\le 1$. In order to prove the lower bound, we may assume w.l.o.g.~that $\vert\langle M_1M_2\rangle\vert\le 1/2$, in which case $\vert 1-\langle M_1M_2\rangle\vert\sim 1$. Now, the numerator in the rhs.~of the $M$-resolvent identity \eqref{eq:identity} is of order one, since $1\gtrsim \vert\langle M_1\rangle-\langle M_2\rangle\vert\gtrsim \vert\langle\Im M_1\rangle\vert + \vert\langle\Im M_2\rangle\vert\gtrsim 1$. 
	Thus, by \eqref{eq:identity} again, we find that $\left\vert (z_1-z_2)\langle M_1M_2\rangle - \langle M_1(D_1-D_2)M_2\rangle\right\vert \sim 1$,
	so, in particular,
	\begin{equation*}
		1\lesssim \left\vert (z_1-z_2)\langle M_1M_2\rangle - \langle M_1(D_1-D_2)M_2\rangle\right\vert\lesssim \vert\langle M_1M_2\rangle\vert + \Delta.
	\end{equation*}
	Therefore, for some constant $c>0$ which depends only on $L$ and $\kappa$ we have
	\begin{equation*}
		\vert\langle M_1M_2\rangle\vert \ge c-\Delta\gtrsim 1, 
	\end{equation*}
	i.e. we get the desired lower bound for $\vert\langle M_1M_2\rangle\vert$.
	\\[2mm]
	\underline{Part (2):} Assume w.l.o.g.~that $\rho_1(z_1) \ge \kappa/2$. The derivative $\partial_{z_1}\mathfrak{s}(z_1,z_2)$ can be computed explicitly as
	\begin{equation*}
		\partial_{z_1}\mathfrak{s}(z_1,z_2) = \frac{\langle M_1^2(D_1-D_2)M_2\rangle \langle M_1M_2\rangle -\langle M_1(D_1-D_2)M_2\rangle\langle M_1^2M_2\rangle}{\langle M_1M_2\rangle^2(1-\langle M_1^2\rangle)}
	\end{equation*}
	and we note that, by analogous reasoning as in part (1), the numerator is bounded from above by $\Delta$. Since $|\langle M_1M_2\rangle|\sim 1$, from part (1), it holds that
	\begin{equation*}
		\vert\partial_{z_1}\mathfrak{s}(z_1,z_2)\vert\lesssim \frac{\Delta}{\vert 1-\langle M_1^2\rangle\vert}\lesssim \Delta,
	\end{equation*}
	where in the last step we used the bound $\vert 1-\langle M_1^2\rangle\vert\gtrsim \rho_1(z_1)^2$ with the aid of Proposition \ref{prop:stab}.
\end{proof}

\begin{proof}[Proof of Lemma \ref{lem:f}]
	The argument is split in two parts: First, we prove existence and uniqueness of the energy renormalization function  $f$. Second, we estimate the partial derivatives \eqref{eq:deriv_bounds} of $f$ and the renormalized (one point) shift $\mathfrak{s}_0$. 
	\\[2mm]
	\underline{Part (1): Existence and uniqueness of $f$.} First, from Lemma \ref{lem:f}, we have that, for $z_1$ with $|z_1| \le L$ and $\Im z_1 < 0$, it holds that $\vert\ms(z_1,z_2)\vert\le C\Delta$ for some $C > 0$. For fixed $z_2 = E_2 + \ii \eta_2$, we introduce the auxiliary (differentiable) function 
	\begin{equation*}
		h(E_1):=E_1-E_2-\Re \ms (E_1-\ii\eta_1,E_2+\ii\eta_2)\,, 
	\end{equation*}
	which has the property that $h(E_1)<0$ for $E_1< E_2-C\Delta$, and $h(E_2)>0$ for $E_1>E_2+C\Delta$. Hence, $h(E_1)=0$ has a solution in $\mathcal{I} := [E_2-C\Delta,E_2+C\Delta]$. To see uniqueness, we differentiate $h$ and find that $h'(E_1) \ge 1 - c \Delta$ for $E_1\in \mathcal{I}$ and some $c > 0$ by means of \eqref{eq:s_deriv_bound} from Lemma \ref{lem:gen_shift}.  Thus $h$ has a unique zero on $\mathcal{I}$ (and hence in $(-L,L)$) which we denote by $f(E_2) = f^{\eta_1, \eta_2}(E_2)$ -- the desired energy renormalization function. Differentiability of $f$ easily follows from the implicit function theorem.
	\\[2mm]
		\underline{Part (2): Bounds on derivatives.} Differentiating the identity $h(f^{\eta_1, \eta_2}(E_2))=0$ in $E_2$, we find that 
		\begin{equation*}
			\partial_{E_2} f^{\eta_1, \eta_2}(E_2) = \frac{1+\Re \partial_2 \ms(f(E_2)-\ii\eta_1,E_2+\ii \eta_2)}{1-\Re\partial_1\ms (f(E_2)-\ii\eta_1,E_2+\ii \eta_2))} = 1 +\mathcal{O}(\Delta),
		\end{equation*}
		by means of  \eqref{eq:s_deriv_bound} from Lemma \ref{lem:gen_shift}. Here, $\partial_j \mathfrak{s}$ denotes the partial derivative of $\mathfrak{s}$ w.r.t.~its $j^{\rm th}$ argument. 
		Similarly, 
		\begin{equation*}
			\partial_{\eta_1}f^{\eta_1,\eta_2}(E_2)=-\frac{\Re\left[\ii\partial_1\ms\right]}{1-\Re \left[ \partial_1\ms\right]},\quad \partial_{\eta_2}f^{\eta_1,\eta_2}(E_2)=\frac{\Re\left[\ii \partial_1\ms\right]}{1-\Re \left[ \partial_2\ms\right]}\,, 
		\end{equation*}
	where $\ms$ has arguments $f(E_2)-\ii \eta_1$ and $E_2+\ii \eta_2$. This concludes the bound $\left\vert \partial_{\eta_j} f^{\eta_1,\eta_2}(E_2)\right\vert\lesssim \Delta$ for $j=1,2$. The bound on $\vert\partial_{E_2}\ms_0(E_2)\vert$ is obtained in a similar fashion and thus left to the reader. 	
\end{proof}

\begin{proof}[Proof of Lemma \ref{lem:Im}] The proof is divided in two parts: In the first part, we prove \eqref{eq:Ims} for $\eta_1=\eta_2=+0$. In the second part of the argument, we treat the the general case as a perturbation thereof. 
\\[2mm]
	\underline{Part (1): Proof on the real line.} Applying the $M$-resolvent identity \eqref{eq:identity} for $z_1:=f(E)-\ii 0$ and $z_2:=E+\ii0$ and using Proposition \ref{prop:stab}, we find that 
	\begin{equation*}
		\left\vert\frac{\langle M_1(z_1)\rangle - \langle M_2(z_2)\rangle}{f(E)-E-\ms_0(E)}\right\vert\lesssim\frac{1}{\Delta^2}\,. 
	\end{equation*}
	Since the numerator on the lhs.~is of order one and the real part of the denominator vanishes by definition of $f(E)$, we deduce that
	\begin{equation*}
		\Delta^2\lesssim \left\vert\Im\left[f(E)-E-\ms_0(E)\right]\right\vert = \vert \Im\ms_0(E)\vert\,,
	\end{equation*}
	i.e.~we have a lower bound on the \emph{modulus} of $ \Im\ms_0(E)$. To turn this into a lower bound on  $\Im\ms_0(E)$ itself, we need to show that it is positive. 
	
This will be done via a proof by contraction: Suppose that $\Im \ms_0(E)<0$. By \eqref{eq:identity} for $z_1:=f(E)-\ii0$, $z_2:=E+\ii0$ we get
\begin{equation}
	\frac{\langle M_1M_2\rangle}{1-\langle M_1M_2\rangle} = \frac{\langle M_1\rangle - \langle M_2\rangle}{-\ii\Im\ms_0(E)}\,. 
	\label{eq:id_on_Re_line}
\end{equation}
Since $\Im\left[\langle M_1\rangle - \langle M_2\rangle\right] = -c$ for some $c > 0$ and $	|\Re\left[\langle M_1\rangle - \langle M_2\rangle\right]| \lesssim \Delta$, we obtain, using our assumption $\Im \ms_0(E)<0$, 
\begin{equation*}
	\langle M_1\rangle - \langle M_2\rangle = \vert \langle M_1\rangle-\langle M_2\rangle\vert \ee^{-\frac{\ii\pi}{2}+\ii\mathcal{O}(\Delta)}\quad \text{and}\quad \frac{\langle M_1\rangle - \langle M_2\rangle}{-\ii\Im\ms_0(E)}= \left\vert \frac{\langle M_1\rangle - \langle M_2\rangle}{-\ii\Im\ms_0(E)}\right\vert \ee^{\ii\pi +\ii\mathcal{O}(\Delta)}\,, 
\end{equation*}
where here and in the following $\mathcal{O}(\Delta)$ is real-valued. In a similar way, we find that	$\langle M_1M_2\rangle=1+\mathcal{O}(\Delta) + \ii \mathcal{O}(\Delta)$ and $\langle M_1M_2\rangle = \vert\langle M_1M_2\rangle\vert \ee^{\ii\mathcal{O}(\Delta)}$. Hence, \eqref{eq:id_on_Re_line} implies
\begin{equation*}
	1-\langle M_1M_2\rangle = \left\vert \frac{-\ii\Im\ms_0(E)}{\langle M_1\rangle - \langle M_2\rangle}\langle M_1M_2\rangle\right\vert \ee^{\ii\pi + \mathcal{O}(\Delta)}\,, 
\end{equation*}
i.e., in particular, $\Re \left[ 1-\langle M_1M_2\rangle\right]<0$. On the other hand, it holds that $\Re \left[ 1-\langle M_1M_2\rangle\right] \ge 1-\vert \langle M_1M_2\rangle\vert\ge 0$, so we arrived at a \emph{contradiction} and thus $\Im \mathfrak{s}_0(E) > 0$ and 	$\Im\ms_0(E)\gtrsim \Delta^2$.

For part (1), we are now left to prove $\vert\Im\ms_0(E)\vert\lesssim \Delta^2$, which is done via a perturbative argument in Appendix \ref{app:technical}. This concludes part (1), i.e.~$\Im \mathfrak{s}_0^{0,0}(E) \sim \Delta^2$. 
\\[2mm]
\underline{Part (2): Extension away from the real line.} By \eqref{eq:deriv_bounds} and the fundamental theorem of calculus, we have
\begin{equation*}
	\left\vert \Im\ms_0^{\eta_1,\eta_2}(E)-\Im \ms_0^{0,0}(E)\right\vert \le \left\vert\int\limits_0^{\eta_1} \partial_{\zeta_1} \ms_0^{\zeta_1,\eta_2}(E)\dif\zeta_1\right\vert + \left\vert\int\limits_0^{\eta_2} \partial_{\zeta_2} \ms_0^{0,\zeta_2}(E)\dif\zeta_2\right\vert \lesssim \Delta(\eta_1+\eta_2)\,. 
\end{equation*}
Hence, if $0 <\eta_j \le c_2 \Delta$ for some $c_2 > 0$ small enough, we obtain $\Im\ms_0^{\eta_1,\eta_2}(E)\sim\Im\ms_0^{0,0}(E)\sim \Delta^2$.
\end{proof}

\section{Contour integration: Proof of technical lemmas from Section \ref{subsec:contint}} \label{sec:conintproof}

The goal of this section is to give the proofs of the technical lemmas from Section \ref{subsec:contint}, for which we recall the construction of the contours $\gamma_1, \gamma_2$ from Section \ref{sec:pfecho1long}, in particular \eqref{eq:contdec1}--\eqref{eq:contdec2} and Figure \ref{fig:1}, and the definition of the $[a,b]$ interval from the beginning of Section \ref{subsec:contint}. 

In all of the estimates below, we will frequently use the following simple tools: 
\begin{itemize}
\item To gain $1/t$-factors from the oscillatory $\ee^{\ii t (z_1 -z_2)}$, we integrate by parts.
\item When pulling absolute values inside an integral, we bound $|\ee^{\ii t (z_1 -z_2)}| \lesssim 1$ (recall $|\Im z_j| \lesssim 1/t$). 
\item The convolution of two Cauchy kernels yields another Cauchy kernel: For $\eta_j > 0$ and $E_j \in \R$, $j\in [2]$ it holds that 
\begin{equation} \label{eq:CKconv}
\int_\R \frac{\eta_1}{(x-E_1)^2 + \eta_1^2} \frac{\eta_2}{(x-E_2)^2 + \eta_2^2} \dif x \lesssim \frac{\eta_1+\eta_2}{(E_1-E_2)^2 + (\eta_1 + \eta_2)^2} \,. 
\end{equation}
\end{itemize}
We now turn to the proofs of the lemmas from Section \ref{subsec:contint}.

\subsection{The second line of \eqref{eq:I_final} is negligible: Proof of Lemma \ref{lem:2nd_term}}
We discuss the contributions from the flat and semicircular part of $\gamma_2$ separately (recall \eqref{eq:contdec2}). 

First, the smallness of the integral over $\gamma_2^{(2)}$ (the semicircular part) is granted by the factor $\ee^{t\Im z_2}$ (note that $\Im z_2 \in [-R + \eta_2, \eta_2]$) and the estimate $|\langle M(E_0+i\eta_0,z_2)\rangle|\lesssim 1$, which follows from \eqref{eq:stab_bound}. More precisely, we have that
\begin{equation}
	\left\vert \oint_{\gamma_2^{(2)}} \ee^{\ii t(E_0+\ii \eta_0-z_2)}\left\langle M(E_0+\ii \eta_0,z_2)\right\rangle\dif z_2\right\vert \lesssim R\int_{\pi}^{2\pi} \ee^{tR\sin\theta}\dif\theta\lesssim \frac{1}{t}\,. 
	\label{eq:arc_bound}
\end{equation}

Next, we bound the integral over $\gamma_2^{(1)}$ -- the flat part. As a first step, integration by parts yields
\begin{equation*}
	\left\vert\int_{\gamma_2^{(1)}} \ee^{\ii t(E_0+\ii\eta_0-z_2)}\left\langle M(E_0+\ii\eta_0,z_2)\right\rangle\dif z_2\right\vert \lesssim \frac{1}{t} +\left\vert\frac{1}{\ii t}\int_{-R}^R \ee^{-\ii tE_2}\partial_{E_2}\left\langle M(E_0+\ii \eta_0,E_2+\ii \eta_2)\right\rangle\dif E_2\right\vert.
\end{equation*}
The derivative can be explicitly computed as
\begin{equation}
	\partial_{z_2}\langle M(z_1,z_2)\rangle =\frac{\langle M_1M_2^2\rangle}{(1-\langle M_2^2\rangle)(1-\langle M_1M_2\rangle)^2}.
	\label{eq:2M_deriv}
\end{equation}
Since $E_0$ is in the bulk of $\rho_1$ and $z_0:=E_0+\ii\eta_0$ and $z_2$ are in the same half-plane we infer $\vert 1-\langle M_1M_2\rangle\vert\gtrsim 1$ and thus 
\begin{equation*}
	\left\vert \oint_{\gamma_2^{(1)}} \ee^{\ii t(E_0+i\eta_0-z_2)}\left\langle M(E_0+\ii\eta_0,z_2)\right\rangle\dif z_2\right\vert \lesssim \frac{1}{t} + \frac{1}{t}\int_{-R}^R \left\vert \frac{1}{1-\langle M_2(E_2+\ii\eta_2)^2\rangle}\right\vert \dif E_2.
\end{equation*}
In order to conclude the proof of Lemma \ref{lem:2nd_term}, we finally use that the one-body stability operator $\vert 1-\langle M_2(E_2+\ii\eta_2)^2\rangle\vert^{-1}$ is locally integrable, see Lemma \ref{lem:int_stab} in Appendix \ref{app:technical}. \qed

\subsection{Cutting tails in the first line of \eqref{eq:I_final}: Proof of Lemma \ref{lem:tails}}
For cutting the tails, we focus on the more critical regime, where both parameters are on the horizontal part of the contours, $z_j \in \gamma_j^{(1)}$ for $j \in [2]$ (recall \eqref{eq:contdec1}--\eqref{eq:contdec2}). Indeed, if this is not the case, a simple computation using Proposition \ref{prop:stab} and arguing similarly to \eqref{eq:arc_bound} yields $(1+\eta_0/\Delta)/t\lesssim 1/t$ as an upper bound for the corresponding integrals.

In the critical regime $z_j \in \gamma_j^{(1)}$ for $j \in [2]$ we carry out only the case $E_2 = \Re z_2 \in[b,R]$; for $E_2\in [-R,a]$ the argument is identical. Let $\delta:=(b-E_0)/2$ and split the region of the $E_1 = \Re z_1$-integration into the two parts, $[b-\delta,2R]$ and $[-2R,b-\delta]$. In the first regime, using $|E_1- E_0| \gtrsim 1$ and, from Proposition~\ref{prop:stab},  $|\langle M(E_1-\ii\eta_1,E_2+\ii \eta_2)\rangle| \lesssim \big((E_1-E_2)^2+\Delta^2\big)^{-1}$, we find that
\begin{equation*}
	\begin{split}
		  \int_{b-\delta}^{2R}\int_b^R \left\vert\frac{\eta_0}{(E_1-\ii \eta_1-E_0)^2 +\eta_0^2}\langle M(E_1-\ii\eta_1,E_2+\ii \eta_2)\rangle \right\vert\dif E_1\dif E_2\lesssim  \frac{\eta_0}{\Delta}.
	\end{split}
\end{equation*}
For $E_1\in [-2R,b-\delta]$, by Proposition \ref{prop:stab} again, we have $\vert\langle M(z_1,z_2)\rangle\vert\lesssim 1$, since $\vert E_1-E_2\vert\sim 1$. Using this and integration by parts in $E_2$, similarly to the proof of Lemma \ref{lem:2nd_term}, in combination with \eqref{eq:2M_deriv} and Lemma \ref{lem:int_stab}, we find that
\begin{equation*}
		\left\vert  \int_{-2R}^{b-\delta}\int_b^R \ee^{\ii t(z_1-E_2-\ii \eta_2)}\frac{\eta_0}{(E_1-\ii \eta_1-E_0)^2 +\eta_0^2}\langle M(E_1-\ii \eta_1,E_2+\ii \eta_2)\rangle\dif E_1\dif E_2\right\vert \lesssim \frac{1}{t} \,. 
\end{equation*}
This finishes the proof of Lemma \ref{lem:tails}. \qed

\subsection{First replacement: Proof of Lemma \ref{lem:1st_repl}}
Let $\delta > 0$  be such that $[a-\delta,b+\delta]$ is in the bulk of $\rho_1$. We now compare the two integrals on the lhs.~and rhs.~of \eqref{eq:firstrepl} by taking their difference. Using integration by parts, the contribution from  $(z_1, z_2) \in \gamma_1^{(2)}\times [a,b]$ is bounded by $\eta_0/t$. Analogously to the proof of Lemma~\ref{lem:tails}, we also find that the contribution from $([-R,R]\setminus [a-\delta,b+\delta])\times [a,b]$ is bounded by $\eta_0$, since in this regime $\vert\langle M(z_1,I,z_2)\rangle\vert\lesssim 1$ and  $\vert z_1-z_2-\ms_0(z_2)\vert^{-1}\gtrsim 1$.

We are hence left to estimate the contribution from the region $[a-\delta,b+\delta]\times [a,b]$. Using that $\vert \ms(z_1,z_2)-\ms_0(E_2)\vert \lesssim \Delta \vert E_1-\Re f(z_2)\vert$ by means of Lemma \ref{lem:gen_shift}, we find that this can be bounded by 
\begin{equation*}
		\mathcal{E}:= \int_{a-\delta}^{b+\delta}\int_a^b \frac{\eta_0}{(E_1-E_0)^2+\eta_0^2} \cdot\frac{\Delta \vert E_1-\Re f(z_2)\vert}{\vert z_1-z_2-\ms(z_1,z_2)\vert \cdot\vert z_1-z_2-\ms_0(E_2)\vert}\dif E_1\dif E_2\,. 
\end{equation*}

To have better control on $\mathcal{E}$, we now bound the denominators in the second factor from below. 
First, using the definition of $\mathfrak{d}$ from the formulation of Lemma \ref{lem:1st_repl}, we get
\begin{equation}
	\vert z_1-z_2-\ms_0(E_2)\vert^2 = (E_1-E_2-\Re \ms_0(E_2))^2 + (\eta_1+\eta_2+\Im \ms_0(E_2))^2\gtrsim (E_1-f(E_2))^2+\mathfrak{d}^2\,. \label{eq:denom1}
\end{equation}
Next, using that $\vert z_1-z_2-\ms(z_1,z_2)\vert\gtrsim\Delta^2$, as simple consequence of the stability bound \eqref{eq:stab_bound}, we infer 
\begin{equation}
	\vert z_1-z_2-\ms(z_1,z_2)\vert^2 \sim \vert z_1-z_2-\ms(z_1,z_2)\vert^2 +\Delta^4\gtrsim (E_1-E_2-\Re \ms(z_1,z_2))^2 + \Delta^4\,. 
	\label{eq:denom2}
\end{equation}
Finally, using the defining properties of the renormalization function $f$ given in Lemma \ref{lem:f}, \eqref{eq:s_deriv_bound} from Lemma \ref{lem:gen_shift}, and the triangle inequality, one easily sees that
\begin{equation}
	\vert E_1-E_2-\Re \ms(z_1,z_2)\vert\sim \vert E_1-f(E_2)\vert.
	\label{eq:equiv}
\end{equation}

Hence, combining \eqref{eq:denom1} and \eqref{eq:denom2}--\eqref{eq:equiv} we find that
\begin{equation*}
	\begin{split}
		\mathcal{E} &\lesssim \int_{a-\delta}^{b+\delta}\int_a^b \frac{\eta_0}{(E_1-E_0)^2+\eta_0^2}\cdot\frac{\Delta \vert E_1-f(E_2)\vert}{\vert E_1-f(E_2)\vert^2 +(\min\lbrace \mathfrak{d},\Delta^2\rbrace)^2}\dif E_1\dif E_2\\
		&\lesssim \int_{a-\delta}^{b+\delta}\frac{\eta_0\Delta(\vert\log\Delta\vert+\vert\log\mathfrak{d}\vert)}{(E_1-E)^2+\eta_0^2}\dif E_1 \lesssim \Delta(\vert \log \Delta\vert+\vert\log\mathfrak{d}\vert).
	\end{split}
\end{equation*}
where in the second step we changed the integration variable from $E_2$ to $f(E_2)$ and employed \eqref{eq:deriv_bounds} from Lemma \ref{lem:f}. This concludes the proof of Lemma \ref{lem:1st_repl}. \qed

\subsection{Residue computation after the first replacement: Proof of Lemma \ref{lem:1st_comp}}
Using the integral representation $\langle M_1(z_1)\rangle = \int_{\R}\rho_1(x)(x-z_1)^{-1}\dif x$ and carrying out the residue computation (note that \eqref{eq:1_condition} ensures $z_2+\ms_0(E_2)$ is encircled by the contour $\gamma_1$), we find the lhs.~of \eqref{eq:rescompfirst} to equal
\begin{equation*}
		-\frac{1}{2 \pi \ii} \int_a^b \dif E_2 \int_{\R} \ee^{\ii t(x-E_2 - \ii \eta_2 )}\frac{\eta_0}{(x-E_0)^2+\eta_0^2}\cdot\frac{\rho_1(x)\dif x}{x-(E_2 + \ii \eta_2 +\ms_0(E_2))} + \mathcal{E}_1 + \mathcal{E}_2 \,, 
\end{equation*}
where we introduced the shorthand notations
\begin{equation*}
	\begin{split}
		&\mathcal{E}_1:=- \frac{1}{4\pi }\int_a^b \ee^{\ii t(E_0+\ii \eta_0-E_2-\ii \eta_2)}\frac{\langle M_1(E_0+\ii \eta_0)\rangle - \langle M_2(E_2+\ii \eta_2)\rangle}{E_0+\ii \eta_0-(E_2+\ii \eta_2+\ms_0(E_2))}\dif E_2,\\
		&\mathcal{E}_2:=\frac{1}{2\pi \ii }\int_a^b \ee^{\ii t\ms_0(E_2)} \frac{\eta_0\left(\langle M_1(E_0+\ii \eta_0)\rangle -\langle M_2(E_2+\ii \eta_2)\rangle\right)}{(E_2+\ii \eta_2+\ms_0(E_2)-E_0)^2 +\eta_0^2}\dif E_2\,. 
	\end{split}
\end{equation*}
Moreover, we shall abbreviate $z_0:=E_0+\ii \eta_0$, $z_2:=E_2+\ii \eta_2$. Then, to estimate $\mathcal{E}_1$, we employ integration by parts and find that, since $\left\vert\partial_{E_2}\langle M_2(z_2)\rangle\right\vert\lesssim 1$ as $\rho_2(z_2) \gtrsim 1$, using \eqref{eq:deriv_bounds} from Lemma \ref{lem:f}, and recalling the definition of $\mathfrak{a}$ from the formulation of Lemma \ref{lem:1st_comp}, 
\begin{equation*}
	\left\vert \partial_{E_2}\frac{\langle M_1(z_0)\rangle - \langle M_2(z_2)\rangle}{z_0-(z_2+\ms_0(E_2))}\right\vert \lesssim \frac{1}{\vert E_0- f(E_2)\vert + \mathfrak{a}} + \frac{\vert \langle M_1(z_0)\rangle - \langle M_2(z_2)\rangle\vert}{\vert E_0- f(E_2)\vert^2 + \mathfrak{a}^2}.
\end{equation*}
Applying the $M$-resolvent identity \eqref{eq:identity} to $z_0$ and $z_2$ we infer, by application of the stability bound from Proposition \ref{prop:stab} together with \eqref{eq:s_bound} and $\eta_2 \lesssim 1/t$, $\eta_0 \lesssim \Delta$, that $\vert \langle M_1(z_0)\rangle - \langle M_2(z_2)\rangle\vert \lesssim \vert E_0- f(E_2)\vert + \Delta+ 1/t$, and hence
\begin{equation*}
	\begin{split}
		&\vert\mathcal{E}_1\vert \lesssim \frac{1}{t}+\frac{1}{t}\int_a^b \left(\frac{1}{\vert E_0-f(E_2)\vert + \mathfrak{a}}+\frac{\vert E_0-f(E_2)\vert + \Delta+ t^{-1}}{\vert E_0- f(E_2)\vert^2 + \mathfrak{a}^2}\right)\dif E_2\lesssim \frac{\vert\log\mathfrak{a}\vert}{t}+\frac{\Delta + t^{-1}}{t\mathfrak{a}}.
	\end{split}
\end{equation*}
Similarly, $\mathcal{E}_2$ admits the bound $\vert\mathcal{E}_2\vert\lesssim  \eta_0\vert\log\mathfrak{a}\vert +\eta_0 \mathfrak{a}^{-1} (\Delta + t^{-1})$. 
This finishes the proof of Lemma~\ref{lem:1st_comp}. \qed

\subsection{Second replacement: Proof of Lemma \ref{lem:2nd_repl}}
The argument is split in two parts. First, we estimate the error of the second replacement within the interval $[a,b]$. Then, we put back the tails to complete the full contour integral. 

For the first part, using $	\vert \ms_0(E_2)-\ms_0\vert \lesssim \Delta \vert f(E_2)-E_0\vert$ as a consequence of \eqref{eq:deriv_bounds}, we find the error to be bounded by (a constant times) $\mathcal{E}_1 + \mathcal{E}_2$, where 
\begin{equation}
	\mathcal{E}_1:=\int_{\R} \dif x \int_a^b \frac{\eta_0}{(x-E_0)^2+\eta_0^2}\cdot \frac{\Delta \vert f(E_2)-E_0\vert}{\vert x-(z_2+\ms_0(E_2))\vert^2}\dif E_2
	\label{eq:2nd_repl_E1}
\end{equation}
and $\mathcal{E}_2$ is the same integral as $\mathcal{E}_1$, but with $\ms_0(E_2)$ being replaced by $\ms_0$. Next, convolving Cauchy kernels \eqref{eq:CKconv} in the $x$-variable and using \eqref{eq:deriv_bounds} together with the definition of $\mathfrak{b}$ we arrive at
\begin{equation*}
	\mathcal{E}_1\lesssim \frac{\eta_0+\mb}{\mb}\int_a^b \frac{\Delta \vert f(E_2)-E_0\vert}{(f(E_2)-E_0)^2 + (\eta_0+\mb)^2}\dif E_2\lesssim \frac{\eta_0+\mb}{\mb} \Delta\vert\log (\eta_0+\mb)\vert\,. 
\end{equation*}
For $\mathcal{E}_2$, the argument is similar: We simply replace $f(E_2) - E_0$ in the denominator by $E_0 - E_2 - \Re \mathfrak{s}_0$ and estimate $\vert E_0- f(E_2)\vert\lesssim \vert E_0-E_2-\Re\ms_0\vert$ in the numerator. This shows that the error for the first bound is bounded by $(\eta_0+\mb)\mb^{-1}\Delta\vert\log(\eta_0+\mb)\vert$. 

In the second part, we estimate the tails on the rhs.~of \eqref{eq:2nd_repl}. In the regime when $z_2\in\gamma_2^{(2)}$ we find the bound $1/t$, similarly to \eqref{eq:arc_bound}. If instead $z_2\in\gamma_2^{(1)}\setminus ([a,b]+\ii \eta_2)$, say, $E_2 = \Re z_2 \in [b,R]$ for concreteness, we have that $\vert E_0-E_2-\Re\ms_0\vert\sim 1$, so the singularities in $x$ on the rhs.~of \eqref{eq:2nd_repl} are separated from each other. Now, pick $\delta\sim 1$ such that $[E_0-\delta,E_0+\delta]\subset [a,b]$ and $\vert x-E_2-\Re\ms_0\vert\sim 1$ for any $E_2\in [b,R]$, $x\in [E_0-\delta,E_0+\delta]$. Then, for $\vert x-E_0\vert\ge \delta$, it holds that 
\begin{equation*}
		\left\vert \int_{\vert x-E_0\vert\ge \delta}\dif x\int_b^R e^{it(x-E_2-i\eta_2)}\frac{\eta_0}{(x-E_0)^2+\eta_0^2}\cdot\frac{\rho_1(x)\dif x}{x-(E_2+i\eta_2+\ms_0)}\dif E_2\right\vert \lesssim  \eta_0\vert\log \mb\vert\,, 
\end{equation*}
where, in order to get $\mathfrak{b}$, we employed Lemma \ref{lem:Im} and \eqref{eq:deriv_bounds}. Finally, for  $\vert x-E_0\vert\le \delta$, we employ integration by parts in $E_2$ and use $\vert x-E_2-\Re\ms_0\vert\sim 1$ for any $E_2\in [b,R]$, $x\in [E_0-\delta,E_0+\delta]$ to get 
\begin{equation*}
		\left\vert \int_{E_0-\delta}^{E_0+\delta} \dif x \ee^{\ii t(x-\ii\eta_2)}\frac{\eta_0 \, \rho_1(x)}{(x-E_0)^2+\eta_0^2}\int_b^R \ee^{-\ii tE_2}\frac{\dif E_2}{x-(E_2+i\eta_2+\ms_0)}\right\vert \lesssim\frac{1}{t}\,. 
\end{equation*}
This finishes the justification of the replacement \eqref{eq:2nd_repl} and thus the proof of Lemma \ref{lem:2nd_repl}. \qed

\section{Second echo protocol: Proof of Theorem \ref{thm:echo2}} \label{sec:pfecho23}
The argument for part (i) is very similar to that for the proof of Theorem \ref{thm:echo1}~(i). The only two differences are the following: First, the formerly algebraic cancellations $\langle \widetilde{P}  [H_1, H_2]  \rangle = 0$ below \eqref{eq:parabolic} and \eqref{eq:algcancpara} are replaced by the estimate $|\langle \psi, W \phi \rangle| \prec \Vert \psi \Vert \, \Vert \phi \Vert N^{-1/2}$ for deterministic $\phi, \psi \in \C^N$. This follows by residue calculus and using an isotropic global law for the Wigner matrix $W$ together with the fact that the first moment of the semicircular density vanishes, $\int_\R x \rho_{\mathrm{sc}}(x) \rd x = 0$, by symmetry. More precisely, using $\Vert W \Vert \le 2+ \epsilon$ with very high probability,
\begin{equation} \label{eq:Wbound}
	\begin{split}
		|\langle \psi, W \phi \rangle| &= \left| \frac{1}{2 \pi \ii} \oint_{|z|=3} z \langle \psi, (W-z)^{-1} \phi \rangle \rd z \right| \\
		&\lesssim \Vert \psi \Vert \, \Vert \phi \Vert \left| \frac{1}{2 \pi \ii} \oint_{|z|=3} z m_{\mathrm{sc}}(z) \rd z \right| + \Vert \psi \Vert \, \Vert \phi \Vert \mathcal{O}_\prec(N^{-1/2}) \\
		& \lesssim \Vert \psi \Vert \, \Vert \phi \Vert \left| \int_\R x \rho_{\mathrm{sc}}(x) \rd x \right| + \Vert \psi \Vert \, \Vert \phi \Vert \mathcal{O}_\prec(N^{-1/2}) \prec \Vert \psi \Vert \, \Vert \phi \Vert N^{-1/2}
	\end{split}
\end{equation}
where, to go to the last line, we used the Stieltjes representation $m_{\mathrm{sc}}(z) = \int_\R (x-z)^{-1} \rho_{\mathrm{sc}}(x) \rd x$ and simple residue calculus. 
Second, in the analog of \eqref{eq:D2higher} it suffices to estimate all the $\lambda W$ simply by operator norm, recalling $\Vert W \Vert \le 2+ \epsilon$ with very high probability. The rest of the argument goes along the same lines as in the proof of Theorem \ref{thm:echo1}~(i) with straightforward modifications.  

Part (ii) may be derived from \cite[Theorem~2.4]{pretherm}, but here we give a direct proof relying just on
 the argument given in \cite[Section~3.2.1]{pretherm}.
 First, by means of the single resolvent global law, we have that
\begin{equation} \label{eq:singleGlambda}
	\begin{split}
		\langle \psi_0, \mathrm{e}^{\ii t H_\lambda} \mathrm{e}^{- \ii t H_0} \psi_0 \rangle &= \frac{1}{2 \pi \ii} \oint_{\gamma} \mathrm{e}^{\ii t z}\langle \psi_0, G_\lambda(z) \mathrm{e}^{- \ii t H_0} \psi_0 \rangle \rd z \\
		&=  \frac{1}{2 \pi \ii} \oint_{\gamma} \mathrm{e}^{\ii t z}\langle \psi_0, M_\lambda(z) \mathrm{e}^{- \ii t H_0} \psi_0 \rangle \rd z + \mathcal{O}_\prec \big(C(t, \lambda)/\sqrt{N}\big)
	\end{split}
\end{equation}
for some constant $C(t, \lambda) > 0$ depending only on time $t$ and coupling $\lambda$. Next, we approximate $\langle M_\lambda(z) \rangle \approx \overline{m_0(E_0)}$, leading to 
\begin{equation} \label{eq:Mapprox}
	M_\lambda(z)  \approx \frac{1}{H_0 -z - \lambda^2 \overline{m_0(E_0)}}\,.
\end{equation}
Plugging the approximation \eqref{eq:Mapprox} into \eqref{eq:singleGlambda}, we find 
\begin{equation} \label{eq:rescalc}
	\frac{1}{2 \pi \ii} \oint_{\gamma} \mathrm{e}^{\ii t z}\big\langle \psi_0, \big(H_0 -z - \lambda^2 \overline{m_0(E_0)}\big)^{-1} \mathrm{e}^{- \ii t H_0} \psi_0 \big\rangle \rd z = \mathrm{e}^{- \ii \overline{m_0(E_0)} \lambda^2 t} 
\end{equation}
from simple residue calculus for $\lambda>0$ small enough, using that $|m_0(E_0)| \lesssim 1$ (as follows from $\rho_0$ being $C^{1,1}$ around $E_0$; recall \eqref{eq:admiss_spec}) and $\gamma$ encircles the spectrum of $H_0$. We have thus extracted the main term in \eqref{eq:singleGlambda}, and it remains to estimate the errors resulting from the replacements in \eqref{eq:Mapprox}. 

Denoting the spectral decomposition of $H_0$ by $H_0 = \sum_j \mu_j \ket{\bm u_j} \bra{\bm u_j} $ and using Assumption \ref{ass:state}, we have that 
\begin{equation} \label{eq:regspecdec}
	\frac{1}{2 \pi \ii} \oint_{\gamma} \mathrm{e}^{\ii t z}\langle \psi_0, M_\lambda(z) \mathrm{e}^{- \ii t H_0} \psi_0 \rangle \rd z	 = \sum_{\mu_j \in I_\Delta} \langle \psi_0,  \bm u_j \rangle \langle \bm u_j, \psi_t \rangle \widetilde{\vartheta}(j) \,,
\end{equation}
where we denoted $\psi_t := \mathrm{e}^{-\ii t H_0} \psi_0$ and 
\begin{equation} \label{eq:thetadef}
	\widetilde{\vartheta}(j) := \frac{1}{2 \pi \ii} \oint_{\gamma} \frac{\mathrm{e}^{\ii t z}}{\mu_j - z - \lambda^2 \langle M_\lambda(z) \rangle} \rd z\,.
\end{equation}

The key to approximating \eqref{eq:regspecdec} is the following lemma, the proof of which is identical to that of \cite[Lemma~3.3]{pretherm} and so omitted. 
\begin{lemma}[cf.~Lemma 3.3 in \cite{pretherm}] \label{lem:repl}
	Under the above assumptions and notations, for every $j \in [N]$ such that $\mu_j \in I_\Delta$, denote  $\vartheta(j) := (2 \pi \ii)^{-1} \oint_{\gamma} \mathrm{e}^{\ii t z}(\mu_j - z - \lambda^2 \overline{m_0(E_0)})^{-1} \rd z$.
	Then it holds that
	\begin{equation} \label{eq:replacereg}
		\sup_{\mu_j \in I_\Delta} \left| \widetilde{\vartheta}(j) - \vartheta(j) \right| \lesssim \mathcal{E}
	\end{equation}
	for sufficiently small $\lambda > 0$ and $N$ large enough (dependent on $\lambda$, cf.~\cite[Lemma~A.1]{pretherm}). Here, recalling \eqref{eq:rho0} for the definition of $\epsilon_0 = \epsilon_0(N)$, 
	we denoted
	\begin{equation} \label{eq:Ereg}
		{\mathcal{E}} = \mathcal{E}(\lambda, t, \Delta, N):= \lambda^2 t \, \Delta + \lambda \,  (1 + \lambda^2 t)+ \frac{\lambda}{\Delta}\left(1 + \frac{\lambda}{\Delta}\right)+  \lambda^2 t \, \epsilon_0\,. 
	\end{equation}
\end{lemma}

Therefore, by means of Lemma \ref{lem:repl}, employing a Hölder inequality in \eqref{eq:regspecdec}, and using \eqref{eq:rescalc}, we find that 
\begin{equation} \label{eq:echo2final}
	\frac{1}{2 \pi \ii} \oint_{\gamma} \mathrm{e}^{\ii t z}\langle \psi_0, M_\lambda(z) \mathrm{e}^{- \ii t H_0} \psi_0 \rangle \rd z = \mathrm{e}^{- \ii \overline{m_0(E_0)} \lambda^2 t}  + \mathcal{O}(\mathcal{E})\,. 
\end{equation}
Combining with \eqref{eq:singleGlambda} and taking the absolute value square of \eqref{eq:echo2final}, we arrive at \eqref{eq:echolong2}. This concludes the proof of Theorem \ref{thm:echo2}. \qed

\appendix

\section{Additional proofs} \label{app:technical}

\subsection{Upper bound on the renormalized shift: Perturbation argument for Lemma \ref{lem:Im} }
The goal of this section is to prove the upper bound $|\Im \mathfrak{s}_0^{0,0}(E)| \lesssim \Delta^2$, as claimed at the end of part (1) of the proof of Lemma \ref{lem:Im} in Section \ref{sec:stabopshiftpf}.

This is done via a perturbative calculation, which we carry out in a slightly more general setting: Consider two spectral parameters $z_1=E_1-\ii 0$, $z_2=E_2+\ii 0$, such that $E_j$ is in the bulk of $\rho_j$, $j\in [2]$. Introducing the averaged and relative coordinates
\begin{equation*}
	D:=(D_1+D_2)/2,\quad z:=(E_1+E_2)/2+i0,\quad \Theta:=(D_2-D_1)/2-(E_2-E_1)/2\,,
\end{equation*}
we find that $D_1-z_1=D-z-\Theta$ and $D_2-z_2=D-z+\Theta$. 
Let $M$ be the solution of the MDE with the averaged coordinates, i.e.
\begin{equation*}
	-\frac{1}{M}=z-D+\langle M\rangle.
\end{equation*}
Using the identity $MM^* = \Im M/\langle\Im M\rangle$, it is easy to compute by Taylor expansion
\begin{equation} \label{eq:M2M1}
	M_2M_1= \frac{1}{\langle \Im M\rangle}\left( \Im M + 2\ii \Im \left[\Im M\Theta M \right] +2\ii \Im \left[ \frac{\langle \Theta M^2\rangle}{1-\langle M^2\rangle}\Im M\cdot M \right]+\mathcal{O}( \vert\Theta\vert^2)\right)\,,
\end{equation}
where $\mathcal{O}(|\Theta|^2)$ indicates terms containing at least two $\Theta$'s.
Plugging \eqref{eq:M2M1} in the definition of the shift \eqref{eq:shift}, we find that 
\begin{equation*}
\mathfrak{s}(z_1, z_2) + (z_2 - z_1) = - 2\frac{\langle \Theta\Im M\rangle + 2\ii \langle\Theta\Im \left[\Im M\Theta M \right]\rangle +2\ii \left\langle\Theta\Im \left[ \frac{\langle \Theta M^2\rangle}{1-\langle M^2\rangle}\Im M\cdot M \right]\right\rangle+\mathcal{O}(\langle\vert\Theta\vert^3\rangle)}{\langle\Im M\rangle + 2\ii \langle\Im \left[\Im M\Theta M \right]\rangle +2\ii \left\langle\Im \left[ \frac{\langle \Theta M^2\rangle}{1-\langle M^2\rangle}\Im M\cdot M \right]\right\rangle+\mathcal{O}(\langle\vert\Theta\vert^2\rangle)}.
\end{equation*}
which implies 
\begin{equation*}
	\begin{split}
		\frac{\langle \Im M\rangle^2}{2}[\mathfrak{s}(z_1, z_2) + (z_2 - z_1)] = &- \langle\Theta\Im M\rangle\langle \Im M\rangle -2\ii\langle \Im M\rangle\left\langle \left(\Theta -\frac{\langle\Theta\Im M\rangle}{\langle\Im M\rangle}\right)\Im M\Theta\Im M\right\rangle\\
		& -2\ii\langle \Im M\rangle\left\langle \left(\Theta -\frac{\langle\Theta\Im M\rangle}{\langle\Im M\rangle}\right)\Im \left[ \frac{\langle \Theta M^2\rangle}{1-\langle M^2\rangle}\Im M\cdot M \right]\right\rangle + \mathcal{O}(\langle |\Theta|^3\rangle).
	\end{split}
\end{equation*}
Using $\Im[z_2-z_1] = 0$ and $\Im [\langle \Theta \Im M\rangle \langle \Im M \rangle] = 0$, the imaginary part is given by 
\begin{equation*}
	\frac{\langle \Im M\rangle}{4}\Im \mathfrak{s}(z_1, z_2) = - \left\langle \left(\Theta -\frac{\langle\Theta\Im M\rangle}{\langle\Im M\rangle}\right)\Im M\left(\Theta\Im M+\Im \left[ \frac{\langle \Theta M^2\rangle}{1-\langle M^2\rangle} M \right]\right)\right\rangle  + \mathcal{O}(\left\langle\vert\Theta\vert^3\right\rangle)
\end{equation*}
and hence
\begin{equation*}
	\left\vert\Im \ms(z_1,z_2)\right\vert = \frac{4}{\langle\Im M\rangle}\left\vert \left\langle \left(\Theta -\frac{\langle\Theta\Im M\rangle}{\langle\Im M\rangle}\right)\Im M\left(\Theta\Im M+\Im \left[ \frac{\langle \Theta M^2\rangle}{1-\langle M^2\rangle} M \right]\right)\right\rangle  + \mathcal{O}(\langle  |\Theta|^3\rangle)\right\vert \lesssim \langle|\Theta|^2\rangle\,, 
\end{equation*}
since $\Vert \Theta \Vert \lesssim 1$. 
Specializing to the setting of the Lemma \ref{lem:Im} this result means that
\begin{equation*}
	\vert\Im \ms_0^{0,0}(E)\vert \lesssim \Delta^2 + \vert f(E)-E\vert^2 \lesssim \Delta^2\,, 
\end{equation*} 
where in the last step we used \eqref{eq:s_bound} and Lemma \ref{lem:f}. 
This concludes the proof of the upper bound in part~(1) of Lemma \ref{lem:Im}.

\subsection{The one-body stability operator is locally integrable}
In Section \ref{sec:conintproof} we frequently use that the one-body stability operator is locally integrable. This is the statement of the following Lemma.
\begin{lemma}[Integral of one-body stability operator] \label{lem:int_stab} Fix a (large) positive constant $L$. Uniformly  in $\eta\in [0,1]$ and in $D$ satisfying Assumption \ref{ass:M_bound} with constant $L$ we have
\begin{equation}
\int_{-L}^L \frac{\dif E}{\vert 1-\langle M^2(E+i\eta)\rangle\vert}\lesssim 1.
\label{eq:int_stab}
\end{equation}
\end{lemma}

\begin{proof} With the notation \eqref{eq:scdos}, we use the classification of local minima of $\rho$ from \cite[Theorem 7.1]{shape}. This result addresses the case of a \emph{diagonal} deformation, while $D$ in the formulation of Lemma \ref{lem:int_stab} does not need to be diagonal. Since the deterministic approximation $M(z)$ to the resolvent $(H-z)^{-1}$ of a random matrix $H$ depends only on the first two joint moments of entries of $H$, we have that $M(z)$ in \eqref{eq:int_stab} coincides with the deterministic approximation to $(W_{\mathrm{GUE}}+D-z)^{-1}$, where $W_{\mathrm{GUE}}$ is a GUE matrix. Let $U$ be a unitary diagonalizing $D$, i.e.~$U^*DU=D_0$, where $D_0$ is diagonal. Invariance of GUE under unitary conjugations gives that $\widetilde{M}(z):=U^*M(z)U$ is a deterministic approximation to $(W_{\mathrm{GUE}}+D_0-z)^{-1}$, so $\widetilde{M}(z)$ solves the MDE 
\begin{equation*}
-\widetilde{M}^{-1}(z)=z-D_0+\langle \widetilde{M}(z)\rangle,\quad \Im z\Im \widetilde{M}(z)>0 \quad \text{for} \quad  z\in \C\setminus\R\,.
\end{equation*}	
We remark that $\widetilde{M}$ satisfies the assumptions of \cite[Theorem 7.1]{shape}, since $\mathcal{S}=\langle\cdot\rangle$ is flat and by means of Assumption \ref{ass:M_bound}. 

Thus \cite[Theorem 7.1]{shape} applied to $\widetilde{M}$ together with the observation $\langle \widetilde{M}\rangle=\langle M\rangle$ gives that there exist positive constants $\rho_*>0$ and $\delta_*>0$ dependent only on $L$ such that for any local minimum $\tau_0$ of $\rho$ with $\rho(\tau_0)<\rho_*$ one of the following possibilities holds:
	\begin{subequations}
			\begin{alignat}{3}
				&\rho(\tau_0+\omega)\sim \min\lbrace \Delta^{-1/6}\omega^{1/2},\omega^{1/3}\rbrace\,,  \qquad &&\omega\in [0,\delta_*]\,, \qquad &&\text{(left edge)}\label{eq:l_edge}\\
				&\rho(\tau_0+\omega)\sim \min\lbrace \Delta^{-1/6}\vert\omega\vert^{1/2},\vert\omega\vert^{1/3}\rbrace, \qquad &&\omega\in [-\delta_*,0]\,, \qquad && \text{(right edge)}\label{eq:r_edge}\\
				&\rho(\tau_0+\omega)\sim \vert \omega\vert^{1/3}, \quad &&\omega\in [-\delta_*,\delta_*]\,, \qquad &&\text{(cusp)}\label{eq:cusp}\\
				&\rho(\tau_0+\omega)\sim \Tilde{\rho}+\min\lbrace \Tilde{\rho}^{-5}\omega^2,\vert\omega\vert^{1/3}\rbrace,\qquad &&\omega\in [-\delta_*,\delta_*]\,, \qquad &&\text{(internal minimum)} \label{eq:intern_min}
			\end{alignat}
	\end{subequations}
	where $\Tilde{\rho}\sim \rho(\tau_0)$ in \eqref{eq:intern_min}.  In \eqref{eq:l_edge}, $\Delta:=1$ if $\tau_0$ is an extreme left edge of ${\rm supp}\rho$ and $\Delta$ is the length of the gap between the intervals of support which ends at point $\tau_0$ otherwise  (see also \cite[Lemma 7.16]{shape})\footnote{To be consistent with \cite{shape} we use $\Delta$ to denote the size of the gap inside of the proof of Lemma \ref{lem:int_stab}. This should not lead to any confusion with the rest of the paper, where $\Delta$ is used for the Hilbert-Schmidt norm of $D_1-D_2$.}, for the right edge \eqref{eq:r_edge} $\Delta$ is defined similarly.

As a first preparatory step for \eqref{eq:int_stab}, we give a lower bound for $\vert 1-\langle M^2(E+\ii \eta)\rangle\vert$ in terms of $\rho(E)$. In fact, we will show that uniformly in $E\in [-L,L]$ it holds that
\begin{equation}
\vert 1-\langle M^2(E+\ii \eta)\rangle\vert \gtrsim \rho^2(E)\,. 
\label{eq:stab_monot}
\end{equation}
By Lemma \ref{lem:proofstab} the LHS of \eqref{eq:stab_monot} has a lower bound of order $\rho^2(E+\ii \eta) + \eta/\rho(E+\ii \eta)$. Recall from \cite[Proposition 2.4]{shape} that $\rho$ is $1/3$-H{\"o}lder regular, i.e.~there exists a constant $C_0$ depending only on $L$ such that $\vert \rho(z_1)-\rho(z_2)\vert\le C_0\vert z_1-z_2\vert^{1/3}$ uniformly in $z_1,z_2\in\C$ with $\Im z_1 \Im z_2 > 0$ and $\vert z_j\vert\le 2L$, $j=1,2$. If $\eta\le \rho(E)^3/(2C_0)^3$, then $\rho(E+\ii \eta)\sim \rho(E)$, i.e. \eqref{eq:stab_monot} holds. In the complementary case, $\eta> \rho(E)^3/(2C_0)^3$, we have $\rho(E+\ii \eta)<\rho(E)$ and hence $\eta/\rho(E+\ii \eta)\gtrsim \rho^2(E)$, i.e.~\eqref{eq:stab_monot} again holds.

Now, armed with \eqref{eq:stab_monot}, we are ready to prove \eqref{eq:int_stab}. We split the region of integration into several regimes according to the classification of local minima of $\rho$. For each local minimum $\tau_0$ with $\rho(\tau_0)<\rho_*$ the integration over $\tau_0+[-\delta_*,\delta_*]\cap \mathcal{D}$ will be considered separately. Here, $\mathcal{D}=\R$ for cusps and internal minima, $\mathcal{D}=[0,+\infty)$ for left edges and $(-\infty,0]$ for right edges. The set $\mathcal{D}$ is chosen in such a way that $\tau_0+[-\delta_*,\delta_*]\cap \mathcal{D}$ covers the part, where $\rho$ is positive and small. The complementary regimes, the \emph{bulk regime} (where $\rho \ge \rho_*$) and the \emph{gap regime} (where $\rho = 0$), are treated separately. 
\\[2mm]
\underline{Bulk regime:} It holds that $\rho(E)\ge \rho_*$, hence desired bound on the lhs.~of \eqref{eq:int_stab} in the bulk regime immediately follows from \eqref{eq:stab_monot}.
\\[2mm]
\underline{Gap regime:} Let $\tau_1<\tau_0$ be two edges of ${\rm supp}\rho$ such that $\rho(E)=0$ for any $E\in [\tau_1,\tau_0]$; the cases when either $\tau_1$ is an extreme right edge or $\tau_0$ is an extreme left edge are treated similarly. Since $\partial_z M=M^2(1-\langle M^2\rangle)^{-1}$, we have $\vert 1-\langle M^2\rangle\vert^{-1}\le 1 +\vert \langle M'\rangle\vert$. Together with \eqref{eq:l_edge} and \eqref{eq:r_edge} this gives that
\begin{equation*}
\vert 1-\langle M^2(E+i\eta)\rangle\vert \gtrsim \left(\left(\min \lbrace{\vert E-\tau_1\vert, \vert E-\tau_0\vert\rbrace}\right)^2+\eta^2\right)^{1/3},
\end{equation*}
so the integral of $\vert 1-\langle M^2(E+\ii \eta)\rangle\vert^{-1}$ over $E\in [\tau_1,\tau_0]$ is uniformly bounded in $\eta\in [0,1]$.
\\[2mm]
\underline{Internal minimum with $\rho(\tau_0)<\rho_*$:} Using \eqref{eq:stab_monot} along with \eqref{eq:intern_min} we find that
\begin{equation*}
\int_{-\delta_*}^{\delta_*} \frac{\dif\omega}{\vert 1-\langle M^2(\tau_0+\omega+i\eta)\rangle\vert}\lesssim \int_{-\delta_*}^{\delta_*} \frac{\dif\omega}{\rho^2(\tau_0+\omega)}\lesssim \int_0^{\Tilde{\rho}^3} \frac{\dif\omega}{\Tilde{\rho}^2}+\int_{\Tilde{\rho}^3}^{\delta_*}\frac{\dif\omega}{\omega^{2/3}}\lesssim 1 \,. 
\end{equation*}
\\[2mm]
\underline{Cusp regime:} This works in the exact same way as the internal minimum, using \eqref{eq:cusp} instead of \eqref{eq:intern_min}. 
\\[2mm]
\underline{Edge regime:} Let $\tau_0$ be a left edge of $\rho$, for the right edge the argument is the same. First, \cite[Corollary~5.3]{shape} gives that $\vert 1-\langle M^2(z)\rangle\vert\gtrsim \rho(z)(\vert\sigma(z)\vert+\rho(z))$, where $\sigma(z)$ is a 1/3-H{\"o}lder regular function in $\{ z \in \C: \Im z > 0\}$ (by \cite[Lemma 5.5]{shape}) and $\vert\sigma(\tau_0)\vert\sim \Delta^{1/3}$ (by \cite[Theorem 7.7, Lemma 7.16]{shape}). Therefore, there exists a (small) positive constant $c\sim 1$ such that for all $z$ with $\Re z\in [\tau_0, \tau_0+c\Delta]$ and $\Im z\in [0,c\Delta]$ it holds that $\vert\sigma(z)\vert\sim\Delta^{1/3}$. It is easy to see that the integral of the one-body stability operator over $[\tau_0+c\Delta,\tau_0+\delta_*]$ has an upper bound of order one by means of \eqref{eq:stab_monot} and \eqref{eq:l_edge}. In the complementary regime $[\tau_0,\tau_0+c\Delta]$ we distinguish between two cases (i) $\eta\in [0,c\Delta]$ and (ii) $\eta>c\Delta$. In the first case, note that, by the integral representation $\rho(E+i\eta)=\int_{\R} \dif x \rho(x) \eta/((x-E)^2+\eta^2)$ and \eqref{eq:l_edge}, it holds that $\rho(E+\ii \eta)\gtrsim \rho(E)$ for $E\in [\tau_0,\tau_0+c\Delta]$. Thus
\begin{equation*}
\int_0^{c\Delta}\frac{\dif\omega}{\vert 1-\langle M^2(\tau_0+\omega+i\eta)\rangle\vert}\lesssim \Delta^{1/3}\int_0^{c\Delta}\frac{\dif\omega}{\omega^{1/2}(\Delta^{1/2}+\omega^{1/2})}\lesssim 1.
\end{equation*}
In the second case, $\eta>c\Delta$, we use \eqref{eq:stab_monot} and the bound $\vert 1-\langle M^2(z)\rangle\vert\gtrsim |\Im z|$ to get
\begin{equation*}
\int_{0}^{c\Delta}\frac{\dif \omega}{\vert 1-\langle M^2(\tau_0+\omega+i\eta)\rangle\vert} \lesssim \int_{0}^{c\Delta}\frac{\dif \omega}{\rho^2(\tau_0+\omega)+\Delta}\sim \Delta^{1/3}\int_0^{c\Delta}\frac{\dif\omega}{\omega + \Delta^{4/3}}\lesssim 1\,, 
\end{equation*}
which concludes the proof for the regular edge. 
\\[2mm]
A careful examination of the proof shows that all implicit constants in the inequalities above depend only on $L$.
\end{proof}

\end{document}